\documentclass[twocolumn,aps,prl,showpacs, preprintnumbers,letterpaper,amsmath,amssymb,floatfix,superscriptaddress]{revtex4-2}

\usepackage{physics}
\usepackage{graphicx}
\usepackage[colorlinks,urlcolor=blue,citecolor=blue,linkcolor=blue]{hyperref}

\begin{document}
\title{Expansion dynamics of a shell-shaped Bose-Einstein condensate} 

\author{Fan Jia}
\affiliation{Department of Physics, The Chinese University of Hong Kong, Hong Kong SAR, China}
\author{Zerong Huang}
\affiliation{Department of Physics, The Chinese University of Hong Kong, Hong Kong SAR, China}
\author{Liyuan Qiu}
\affiliation{Department of Physics, The Chinese University of Hong Kong, Hong Kong SAR, China}
\author{Rongzi Zhou}
\affiliation{Department of Physics, The Chinese University of Hong Kong, Hong Kong SAR, China}
\author{Yangqian Yan}
\affiliation{Department of Physics, The Chinese University of Hong Kong, Hong Kong SAR, China}
\affiliation{The Chinese University of Hong Kong Shenzhen Research Institute, Shenzhen, China}
\author{Dajun Wang}
\email{djwang@cuhk.edu.hk}
\affiliation{Department of Physics, The Chinese University of Hong Kong, Hong Kong SAR, China}
\affiliation{The Chinese University of Hong Kong Shenzhen Research Institute, Shenzhen, China}

\date{\today}% It is always \today, today,
             %  but any date may be explicitly specified

\begin{abstract}
Bose-Einstein condensates (BECs) confined on shell-shaped surfaces have been proposed as a platform for exploring many nontrivial quantum phenomena on curved spaces. However, as the shell-shaped trapping potential generated with the conventional radio frequency dressing method is very sensitive to gravity, so far experimental studies of shell BECs can only be performed in micro-gravity environments. Here, we overcome this difficulty and create a shell BEC in the presence of Earth's gravity with immiscible dual-species BECs of sodium and rubidium atoms. After minimizing the displacement between the centers of mass of the two BECs with a magic-wavelength optical dipole trap, the interspecies repulsive interaction ensures the formation of a closed shell of sodium atoms with its center filled by rubidium atoms. Releasing the double BEC together from the trap, we observe explosion of the filled shell accompanied by energy transfer from the inner BEC to the shell BEC. With the inner BEC removed, we obtain a hollow shell BEC which shows self-interference as a manifestation of implosion. Our results pave an alternative way for investigating many of the intriguing physics offered by shell BECs. 
\end{abstract}

\maketitle

Introduction---Engineering the external trapping potential to emulate Hamiltonians governing vastly different physical systems is one of the main applications of Bose-Einstein condensates (BEC) of ultracold dilute atomic gases~\cite{Bloch2008}. In the past two decades, BECs in optical lattices~\cite{Greiner2002,Bloch2012,Gross2017}, in lower dimension 2D~\cite{Hadzibabic2011} and 1D traps~\cite{Cazalilla2011}, and in box potentials~\cite{Navon2021} have all been routinely created for investigating a broad range of physics. Another interesting path is creating BEC samples with non-trivial real-space topologies. For example, the persistent flow of toroidal BECs is linked to physics in the SQUID system~\cite{Ramanathan2011,Wright2013}, while its supersonic expansion is analogous to the cosmological expansion of the Universe~\cite{Eckel2018}. BEC trapped on the surface of a sphere, dubbed the shell or bubble BEC, which was first proposed in 2001~\cite{Zobay2001}, is another important case of BEC with non-trivial real-space topology. In recent years, many unique properties of the condensate caused by the topology of the shell structure have been predicted, including distinctive collective modes \cite{Lannert2007,Padavic2017,Sun2018}, interesting thermodynamic behaviors such as expansion induced cooling and depletion of the condensate \cite{Tononi2019,Tononi2020,Rhyno2021}, and the appearance of self-interference patterns due to implosion during free expansion \cite{Lannert2007,Tononi2020}. For rotational velocity above a critical value, the closed shell leads to the formation of stable vortex-antivortex pairs \cite{Padavic2020}. In the thin-shell limit, the shell BEC can also undergo the Berezinskii–Kosterlitz–Thouless (BKT) transition for revealing the connection between BEC and superfluid on curved spaces~\cite{Tononi2019}. 
 
Despite these long-standing interests, experimental investigations on shell BECs are still an effort underway, mainly because of the challenge in its creation. The original proposal for generating the shell-shaped potential relies on the radio-frequency dressed magnetic trap~\cite{Zobay2001,Zobay2004}. Unfortunately, this potential is prone to gravity induced distortion which depletes the density at the top to zero. Because of this, a micro-gravity environment, such as that found in the NASA Cold Atom Laboratory (CAL) on board of the International Space Station, has been assumed as a prerequisite for realizing the shell BEC~\cite{Lundblad2019}. Indeed, very recently, the first shell BEC was successfully created by the CAL project~\cite{Carollo2021}. 
%However, due to technical defects, the shell is still asymmetric~\cite{Carollo2021}.       

\begin{figure*}[t]
\begin{center}
\includegraphics[width = 0.9 \linewidth]{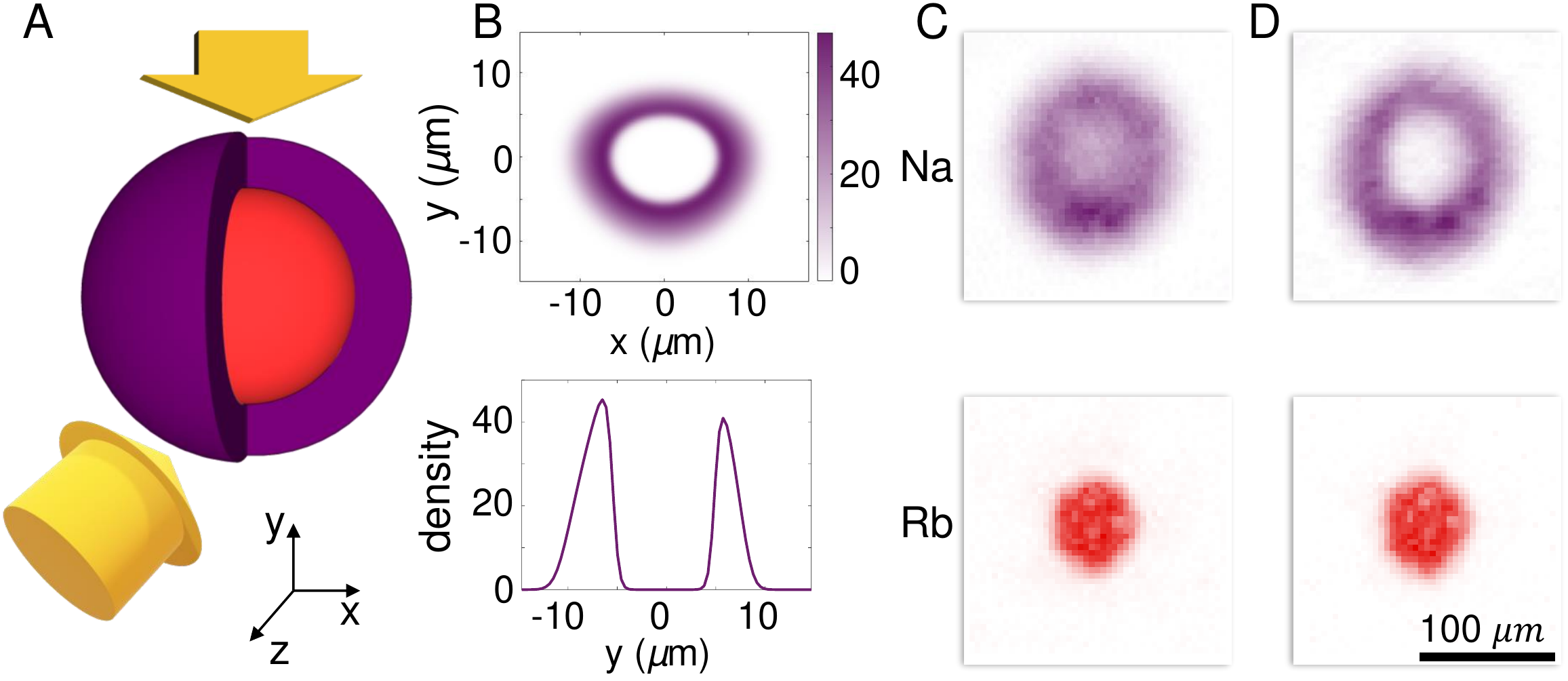}
\end{center}
\caption{Creating a shell-shaped BEC with the immiscible double BEC. (\textbf{A}) Schematic 3D distribution of the $^{23}$Na (purple) and $^{87}$Rb (red) atoms in the magic-wavelength crossed ODT. The probe beams (left-bottom arrow) for absorption imaging propagate along the z-axis. The light sheet optical pumping beam for the shell (top arrow) is introduced from the vertical y-direction. (\textbf{B}) The 2D contour plot of the $^{23}$Na shell in the x-y plane (upper) and its 1D profile along the y-axis (lower) from numerical simulation. In the presence of the y-axis asymmetry due to the gravity induced trapping potential distortion, a closed $^{23}$Na shell can still form. The upper images in (\textbf{C} and \textbf{D}) are OD images of the shell obtained with round and light sheet $^{23}$Na optical pumping beams, respectively. The shell structure is more clearly revealed in the latter configuration. All four images are obtained after co-expansion time $t_{\rm co}=12$ ms . The same color code will be used throughout this paper.}
\label{fig1}
\end{figure*}

We present here the creation and investigation of the shell BEC using a very different method which relies on optically trapped double species BECs of $^{23}$Na and $^{87}$Rb atoms in the immiscible phase~\cite{Ho1996,Pu1998,Trippenbach2000,Lee2016}. We found that $^{23}$Na BECs in a fully closed shell can be formed robustly in the presence of Earth's gravity. This has allowed us to study the expansion dynamics of the shell BEC together with the center BEC or by itself. In the latter case, the implosion of the shell, which manifests as highly repeatable self-interference patterns, is observed~\cite{Lannert2007,Tononi2020}.

In the 1990s, Ho et al.~\cite{Ho1996} and Pu et al.~\cite{Pu1998} have already pointed out that in the immiscible phase of the double BEC of $^{23}$Na and $^{87}$Rb atoms, the $^{23}$Na BEC could form a shell surrounding the $^{87}$Rb BEC. Recently, Wolf et al. revisited this idea~\cite{Wolf2021}. In 2013, we produced the first double BEC of $^{23}$Na and $^{87}$Rb and confirmed that they are indeed immiscible near zero magnetic field~\cite{Wang2013,Wang2015}. Originally, the two condensates were co-trapped in an optical dipole trap (ODT) formed by crossing two 1070 nm laser beams. In the harmonic trap approximation, at this wavelength, the trap oscillation frequencies $\omega$ are not the same. In the vertical (y) direction, this leads to a displacement between the centers of mass of the two BECs due to their different gravitational sags $-g/\omega^2$, where $g$ is the gravitational acceleration. As a result, the $^{23}$Na BEC has a crescent shape with a bottom notch repelled by the $^{87}$Rb atoms~\cite{Wang2015}. Here, we solve this problem by changing the crossed ODT to the ``magic'' wavelength of 946 nm where the two species have the same $\omega$~\cite{Safronova2006}. They thus have the same amount of gravitational sags and nearly overlapping centers of mass.        

In this magic-wavelength trap, the number density ratio $n_2/n_1$ for two independent condensates is proportional to $(U_{11}m_2/U_{22}m_1)^{3/5}$, where $U_{ij}=2\pi \hbar^2 a_{ij}/\mu_{ij}$ are the interaction constants, $m_i$ the atomic masses, and $\mu_{ij}=m_i m_j/(m_i+m_j)$ the reduced masses (with $i=1$ and 2 for $^{23}$Na and $^{87}$Rb, respectively). We prepare $^{23}$Na($^{87}$Rb) in the $\ket{F=1, m_F = -1}$ hyperfine Zeeman state of which the $s$-wave scattering length is $a_{11} = 54.6 a_0$~\cite{Knoop2011} ($a_{22}=100.1 a_0$~\cite{Kempen2002}). With $a_{12} = 76.3a_0$ in low magnetic field~\cite{Guo2022}, the $^{23}$Na-$^{87}$Rb double BEC is in the immiscible phase with $U_{12}>\sqrt{U_{11}U_{22}}$. Since $n_2/n_1=3.4$ for equal number of atoms, the lower density $^{23}$Na BEC will be pushed outward to further lower its density and reduce its overlap with $^{87}$Rb atoms to minimize the total interaction energy. The result of this buoyancy effect is a shell of $^{23}$Na BEC filled with $^{87}$Rb atoms, as illustrated schematically in Fig.~\ref{fig1}A.

The more detailed density distribution of the double BEC can be obtained by numerical simulation~\cite{Note1} of the coupled Gross-Pitaevskii equations (GPEs)~\cite{Antoine2014}
\begin{equation}
i\hbar \frac{\partial \psi_{i}}{\partial t} =[-\frac{\hbar^2 \nabla^2}{2m_{i}} +V_{i}+U_{ii} n_i+U_{ij} n_j]\psi_{i},
\label{eq1}
\end{equation}
where $V_{i}$ are the external trap potentials including contributions from both the ODT and gravity. 
After the evaporative cooling for creating the double BEC, although the centers of mass are nearly together, $V_{i}$ are still severely distorted by gravity. The result of this complication is reflected in the 2D contour plot of the calculated in-trap density of the $^{23}$Na BEC in the x-y plane as presented in the upper panel of Fig.~\ref{fig1}B which indicates that a closed, but not perfectly symmetrical shell is indeed formed. It becomes even more obvious in the 1D density distribution sliced along the vertical direction. For example, the lower panel of Fig.~\ref{fig1}B shows that more atoms are located at the bottom than at the top part of the shell. Nevertheless, our simulation shows that, for a large range of atom number ratios, fully closed $^{23}$Na BEC shells can still be formed readily.

%\red{This is a result of the distorted trap potential and can be understood intuitively by looking at the effective trapping potential experienced by the $^{23}$Na BEC $V_1 + U_{12} n_2$.}

Detection of the shell structure---Our experiment starts from double BECs prepared in the crossed-beam magic-wavelength ODT with typically 1.3$\times 10^5$ $^{23}$Na atoms and 1.2$\times 10^5$ $^{87}$Rb atoms. At the final stage of the evaporative cooling, the measured trap oscillation frequency is $2\pi \times 108$ s$^{-1}$ ($2\pi \times 85$ s$^{-1}$) in the vertical (radial) direction. To detect the atoms, first, we abruptly switching off the trapping laser beams followed by time-of-flight (TOF) expansion. Second, atoms are transferred from the $F = 1$ hyperfine states to the $F = 2$ states by an optical pumping beam. For $^{87}$Rb, the state transfer is done with either an optical pumping beam or a 6.8 GHz microwave. Finally, we detect the atoms by absorption imaging using probe beams on the $F=2 \rightarrow F'=3$ cycling transitions.  
Thanks to the drastically different transition wavelengths, we not only achieve species-selective detection in nearly the same instant but also can study the expansion dynamics of one of the species by removing the other one. Hereafter, we will use $t_{\rm co}$ to denote the expansion time of the filled shell (or in other words, co-expansion time of the double BEC), and $t_{\rm s}^{\rm Na}$ ($t_{\rm s}^{\rm Rb}$) to represent the additional expansion time of the $^{23}$Na($^{87}$Rb) BEC after the $^{87}$Rb($^{23}$Na) BEC is removed. The total TOF after shutting off the ODT is thus $t_{\rm tof} = t_{\rm co} + t_{\rm s}^{\rm Na/Rb}$.

\begin{figure}[t]
\begin{center}
\includegraphics[width = 0.9 \linewidth]{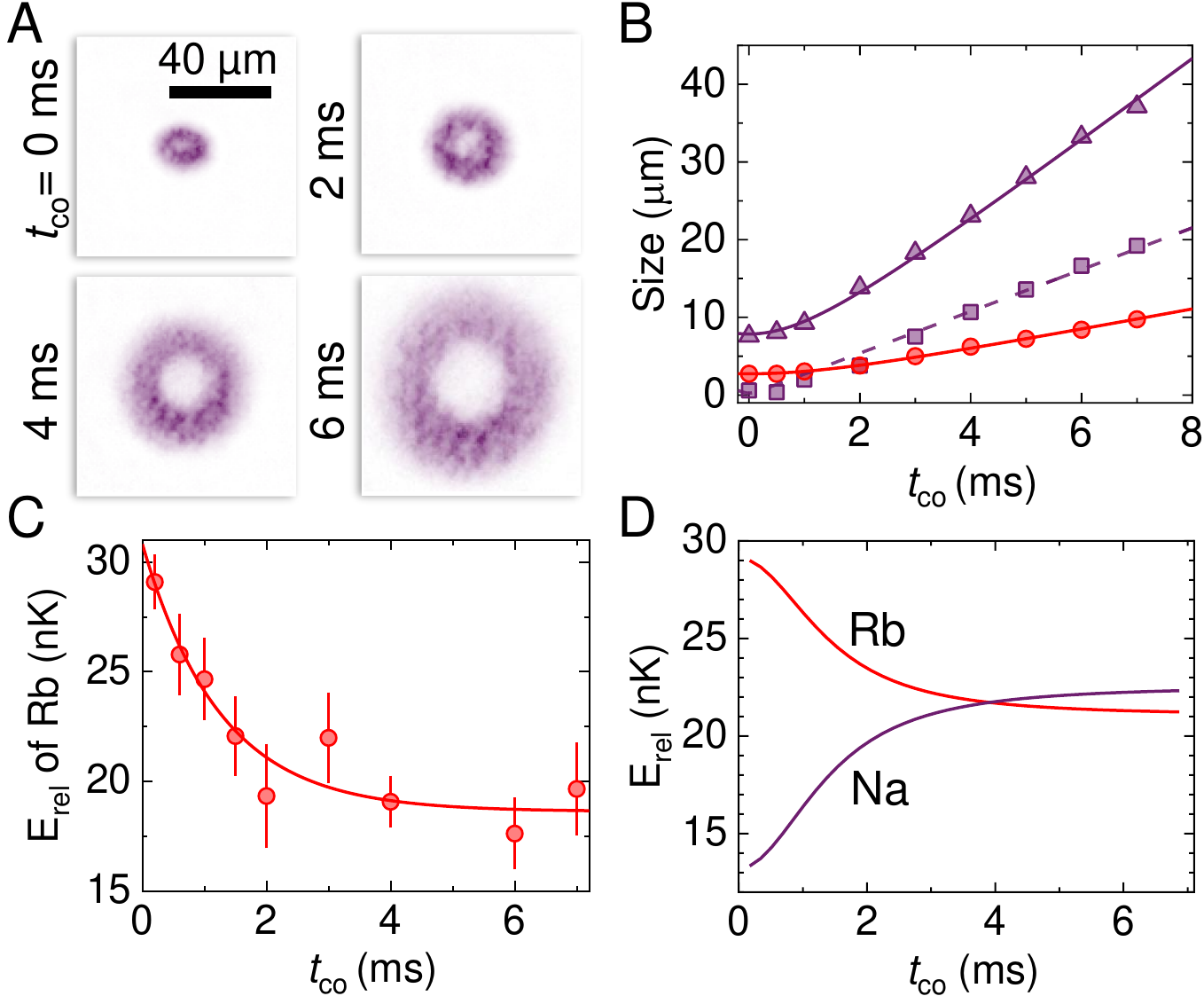}
\end{center}
\caption{Expansion dynamics of the $^{23}$Na shell filled with the $^{87}$Rb BEC. (\textit{A} and \textit{B}) OD images and the extracted inner (squares) and outer (triangles) sizes of the $^{23}$Na shell versus co-expansion time $t_{\rm co}$. For the first 2 ms, the size of the $^{87}$Rb BEC (filled circles) is nearly the same as the inner size of the shell. (\textit{C}) $E_{\rm rel}$ of the $^{87}$Rb BEC after co-expansion with the shell. The solid curve is fit to the shifted exponential decay with a time constant of 1.3(4) ms. (\textit{D}) The numerical simulation confirms the energy transfer from the inner $^{87}$Rb BEC to the $^{23}$Na shell. 
}
\label{fig2}
\end{figure}

Figure~\ref{fig1}C shows the absorption images of the $^{23}$Na and $^{87}$Rb BECs after $t_{\rm co}$ of 12 ms. Due to the integration along the probing direction, the shell structure of the $^{23}$Na BEC manifests only as a lower optical density (OD) in the center. To obtain more detailed information of the shell, we change the $^{23}$Na optical pumping beam to a light sheet~\cite{Andrews1997,Note1} of 15 $\mu$m thick and 800 $\mu$m wide. In this configuration, only a thin slice of the $^{23}$Na atoms in the center part of the shell can be detected by the probe beam. The $^{23}$Na shell thus shows up as a ring in the image as can be seen in Fig.~\ref{fig1}D. However, as shown in Fig.~\ref{fig2}A, since the thickness of the light sheet is limited, to resolve the hollow $^{23}$Na shell, $t_{\rm tof}$ needs to be longer than 2 ms, when the inner radius of the shell becomes larger following the expansion.
For the best performance, the focus of light sheet is adjusted to follow the shell at different $t_{\rm tof}$. 

%For the $^{87}$Rb BEC, a round-shaped optical pumping beam is always used.
    
%A series of absorption images for TOF from 1 ms to 20 ms obtained with this method are shown in Fig.~\ref{fig2}(b). \red{\sout{Each picture is averaged over 12 images.}}  

\textit{Expansion dynamics of the filled shell---} Due to the repulsive interspecies interaction, the filled $^{23}$Na shell expands outward and its outer and inner sizes increasing with $t_{\rm co}$, as shown in Figs.~\ref{fig2}A and B.  Correspondingly, the outward expansion of the $^{87}$Rb BEC is also affected by the $^{23}$Na shell which, not surprisingly, is different from the typical BEC expansion in free space. To reveal this point, we measure the release energy $E_{\rm rel}$~\cite{Mewes1996,Holland1997} of the $^{87}$Rb BEC after different $t_{\rm co}$. We first let the $^{87}$Rb BEC co-expand with the $^{23}$Na shell and then remove the $^{23}$Na shell in less than 100~$\mu$s. Afterwards, the $^{87}$Rb BEC starts to expand in free space and its $E_{\rm rel}$ is extracted from the expansion velocity after further TOF $t_{\rm s}^{\rm Rb}$. As shown in Fig.~\ref{fig2}C, initially $E_{\rm rel}$ decreases rapidly with $t_{\rm co}$, i.e., the $^{87}$Rb BEC is losing energy during the co-expansion. At longer $t_{\rm co}$, $E_{\rm rel}$ finally levels off. During this whole process, about 40\% of the initial $E_{\rm rel}$ are lost. Empirically, we find that the data is fitted very well to a shifted exponential decay with a time constant $\tau = 1.3(4)$ ms.

Since the total energy of the double BEC is conserved, the lost energy of the $^{87}$Rb BEC can only be absorbed by the $^{23}$Na shell, i.e., the expansion of the filled shell is accompanied by energy transfer between the two species. 
This is in analogous to the energy transfer between coupled oscillators, only here it is irreversible due to the free expansion of the shell. 
The energy transfer rate, which is reflected in the loss rate of $E_{\rm rel}$, is proportional to the interspecies interaction energy. It is thus fastest at shorter $t_{\rm co}$ and slows down when the interaction energy is weakened following the expansion. This picture agrees with the observation in Fig.~\ref{fig2}C which is confirmed by numerical simulations of the coupled GPEs plotted in Fig.~\ref{fig2}D. The simulation also confirms that indeed the energy of the $^{23}$Na shell increases during the filled expansion.  

%This can be understood by taking the expansion of the $^{87}$Rb BEC as a confined motion in the potential provided by the interspecies repulsive interaction. The observed energy transfer is then the result of the one-way transformation from the kinetic energy of the $^{87}$Rb atoms to the potential energy which is then converted to the kinetic energy of the $^{23}$Na atoms irreversibly following the shell expansion. 

\begin{figure}[t]
\begin{center}
\includegraphics[width = 0.9 \linewidth]{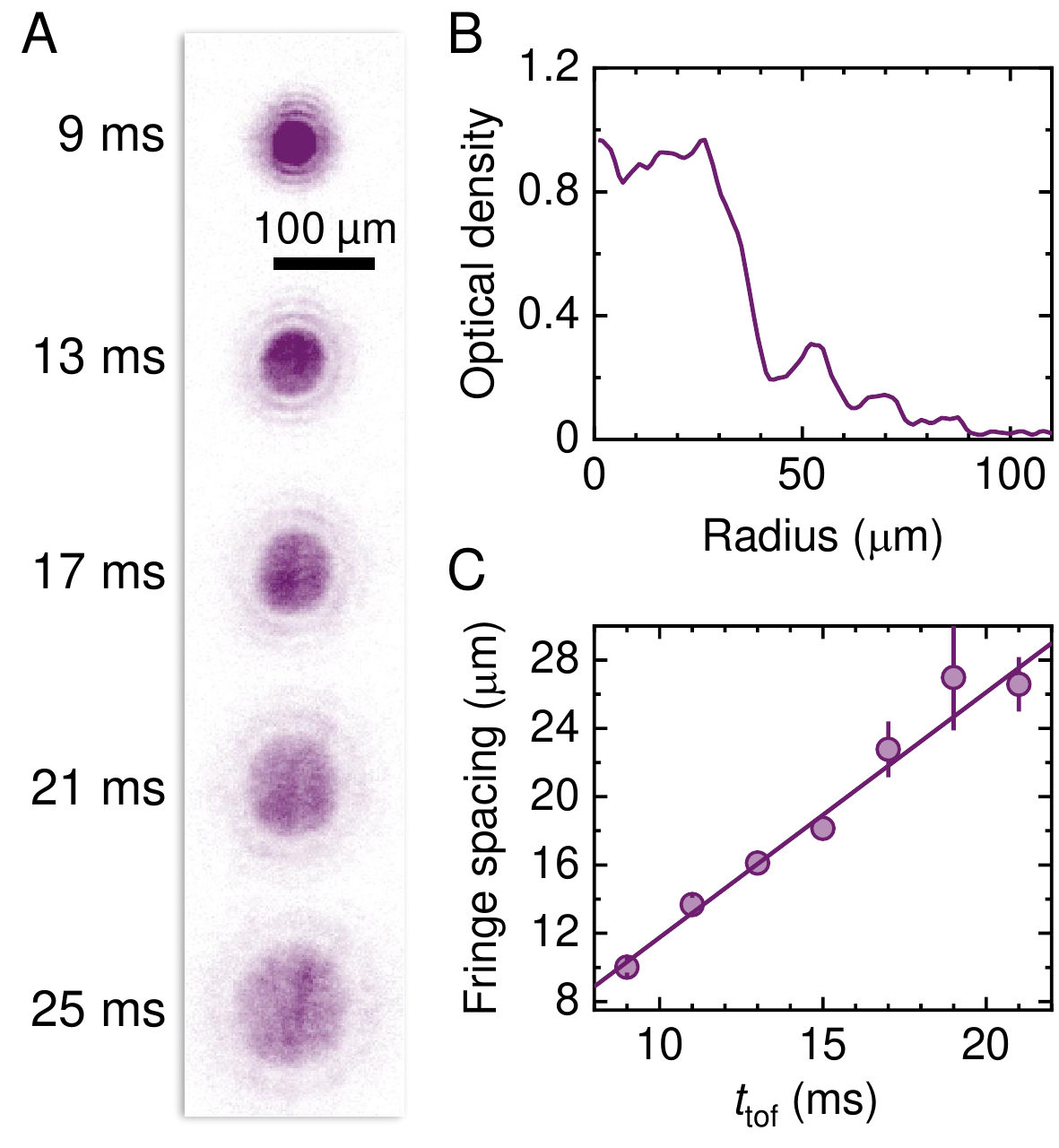}
\end{center}
\caption{Self-interference of the hollow shell BEC. (\textit{A}) OD images at different $t_{\rm tof}$ for the $^{23}$Na shell created after co-expansion time $t_{\rm co}=0.3$ ms. The implosion of the shell leads to self-interference. The contrast of the images is adjusted to enhance the visibility of the fringes. (\textit{B}) Azimuthally averaged profile of the $^{23}$Na shell at $t_{\rm tof}$ = 15 ms. The center peak is flat-top and the contrast of the first fringe is about 30\%. (\textit{C}) The measured fringe spacing versus $t_{\rm tof}$. The solid line is a linear fit to the data.  }
\label{fig3}
\end{figure}

Self-interference of the hollow shell---Next we study the very different expansion dynamics of the hollow $^{23}$Na shell. To this end, after some initial filled expansion time $t_{\rm co}$, we remove the $^{87}$Rb atoms in 100 $\mu$s by pulses of the optical pumping beam and the probing beam. During and after this removal process, we observe no indications that the $^{23}$Na shell is affected. Figure~\ref{fig3}A demonstrates the absorption images at several $t_{\rm tof}$ for the shell created after 0.3 ms of $t_{\rm co}$. Different from the filled expansion, here the $^{23}$Na shell expands both outward and inward. When $^{23}$Na atoms reach the center following the inward implosion, the signature self-interference of the shell BEC~\cite{Lannert2007,Tononi2020} is observed. Figure~\ref{fig3}B shows the partially azimuthally averaged OD~\cite{Note1} versus distance from the shell center for $t_{\rm tof} = 15$~ms. Two important features can be clearly identified. First, the majority of the $^{23}$Na atoms accumulate at the center and form a broad peak with a nearly flat-top high density distribution. Second, concentric interference fringes with much lower densities appear. Both features are in full agreement with theoretical predictions~\cite{Lannert2007,Tononi2020,Wolf2021}. Typically, the first fringe shows up as a nearly fully closed ring while fragments of the third fringe are still visible. The measured contrasts of the fringes are up to 30\%. Considering the light sheet optical pumping beam and the symmetry of the shell, it is conceivable that the first fringe forms a nearly spherical and probably closed shell. For the higher orders fringes, only broken shells are formed, most probably due to the imperfect spherical symmetry of the shell. 

From shot to shot, the main features of the self-interference patterns are highly repeatable because of the fixed relative phase in the initial wave function. This is in stark contrast to the interference between two completely independent BECs in which the fringes vary from shot to shot due to the separated phase evolution of the two BECs~\cite{Andrews1997,Rohrl1997}. However, in both cases, the fringe spacing $\delta$ increases linearly for long TOF. For the shell BEC, we have~\cite{Dalfovo1999,Lannert2007}
\begin{equation}
\delta=\frac{h~t_s^{\rm Na}}{2 r_{0}~m_{\rm Na} },
\label{eq2}
\end{equation}
with $2 r_{0}$ an initial diameter of the shell, and $h$ the Planck constant. Figure~\ref{fig3}C shows $\delta$ extracted from the images in Fig.~\ref{fig3}A. From a linear fit to Eq.~\ref{eq2} with an offset, the fringe spacing increases with a rate of 1.31(4) $\mu$m/ms which is consistent with the simulated value of 1.13 $\mu$m/ms. 
 
% \sout{??? Taking into account the resolution of the imaging system, this agrees well with the measured value.}  

% \blue{Is this the right logical way to discuss this? Do we need to take the resolution into account in the beginning already? How about the simulation?}                

%\sout{To extract $\delta$ from the images, we first enhance the signal by selecting a vertical slice from the central region and obtaining a 1D distribution after a horizontal integration. After masking the central peak, we fit the fringes with a convolution of Gaussian and biased Sine function to get the distance between the first and the second maximum.}

\begin{figure}[t]
\begin{center}
\includegraphics[width = 0.9 \linewidth]{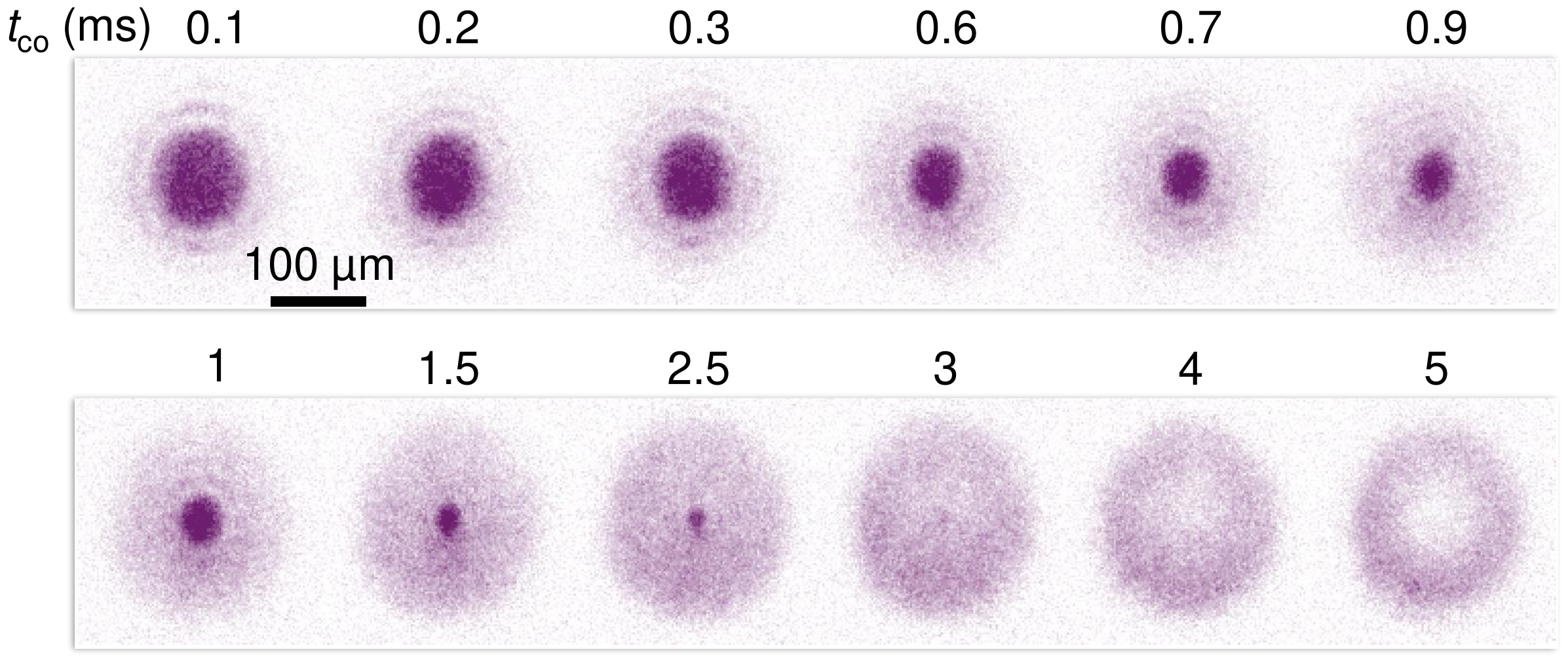}
\end{center}
\caption{Controlling the expansion dynamics of the hollow shell. The self-interference patterns for shells formed after different co-expansion time $t_{\rm co}$ are shown. The total TOF $t_{\rm tof}$ is fixed to 20 ms. The contrast of the images is adjusted to enhance the visibility.}
\label{fig4}
\end{figure}

The self-interference can be tuned by changing the initial $t_{\rm co}$. As shown in Fig.~\ref{fig4} are images of the $^{23}$Na shell with fixed $t_{\rm tof}$ of 20 ms for initially hollow shells created at $t_{\rm co}$ from 0.1 ms to 5 ms. With increasing $t_{\rm co}$, the number of atoms accumulated at the center reduces continuously and vanishes completely at 2.5 ms. For even longer $t_{\rm co}$, no obvious implosion can be observed any more. The visibility of the self-interference fringes also decreases with increasing $t_{\rm co}$ and disappears at around 1 ms. 

These behaviors can be interpreted by the aforementioned energy transfer during the filled expansion. For the first 2 ms, the $^{23}$Na shell gains more and more energy, so both its inner and outer surfaces will expand outward with faster and faster velocities. After the $^{87}$Rb BEC is removed, the motion of the inner surface of the $^{23}$Na shell has to be reversed for implosion to happen. Thus, at the moment when implosion starts, $r_{\rm 0}$ will be larger for longer $t_{\rm co}$. This will result in less accumulation of atoms at the center and, following Eq.~\ref{eq2}, a smaller $\delta$. Eventually, for even longer $t_{\rm co}$, when most of the internal energy of the $^{23}$Na BEC has been converted into kinetic energy of the outward motion, there is simply not enough energy left to reverse the motion for implosion to happen.

\textit{Summary and outlook---} In the current system, the shell BEC is still not in the thin shell limit which is the focus of most theoretical works~\cite{Lannert2007,Padavic2017,Sun2018,Tononi2019,Tononi2020,Rhyno2021}. However, many features of the shell BEC, including the self interference as studied here and the distinctive collective excitation modes~\cite{Lannert2007,Wolf2021,Pu1998}, can still be explored. The double BEC system also provides many additional flexibility. For instance, the in-trap shell thickness can be reduced by increasing the number ratio between the inner BEC and the shell BEC~\cite{Pu1998}. For the $^{23}$Na-$^{87}$Rb system, the previously observed Feshbach resonances~\cite{Wang2013,Guo2022} can be exploited to tune the interspecies interaction. In future investigations, this can be used to control the miscibility to understand the formation dynamics of the shell in a well controlled manner. It is also possible to change the sign of the interspecies interaction for studying the shell implosion in different regimes or even the shell to quantum droplet~\cite{Guo2021} transition.  

%To conclude, we demonstrate a new method for generating  shell BECs in territorial. Instead of shaping the trap potential by rf-dressing, in two-species BECs system, the central repulsive barrier is rather formed by the immersible nature of mixture. We investigate the energy transfer process during explosion in double BECs system, recognize the increasing shell radius results from repulsion of the central core. By removing the core atoms, we observe prominent self-interference fringes on the expanding shell. The unique shell-shape fringes support the evidence of existing hollow shell in our system. Our observation introduces a path toward
%investigating nontrivial topological hollow systems. Our further research will be focused on finding characteristic collective excitation modes[] and the relation between excitation frequencies and shell radius.

\section{Acknowledgments}
We thank Zhichao Guo for the valuable discussions. This work was supported by the Hong Kong RGC General Research Fund (Grants 14301620 and 14301818), the Collaborative Research Fund (Grants C6005-17G, and C6009-20GF) and CUHK Direct Grant No. 4053535.

\renewcommand\thefigure{\thesection S\arabic{figure}}
\renewcommand\theequation{\thesection S\arabic{equation}}
\setcounter{figure}{0}
\setcounter{equation}{0}

\subsection{Preparing the shell BEC in the magic wavelength trap}

\begin{figure}[t]
\begin{center}
\includegraphics[width = 0.9 \linewidth]{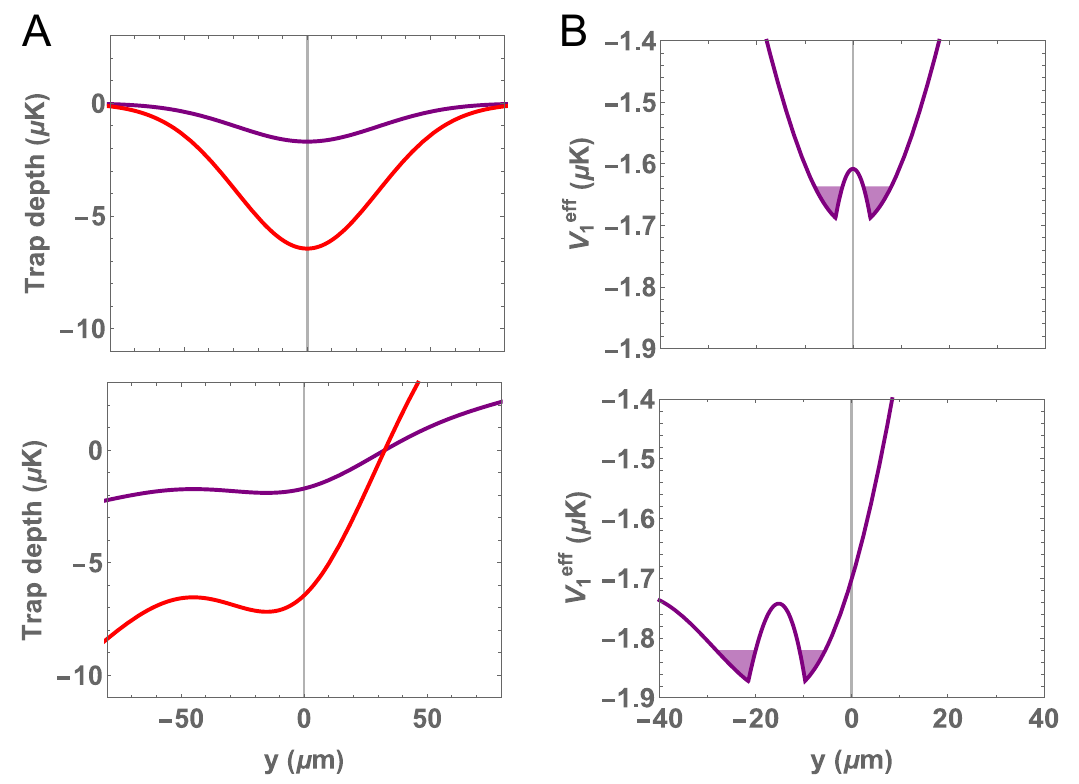}
\end{center}
\caption{The magic wavelength trap and the influence of gravity. (A) The calculated external trapping potentials generated by the 946 nm optical dipole trap for $^{23}$Na and $^{87}$Rb without (top) and with (bottom) gravity. The potentials are shifted and distorted by gravity. The distortion is especially severe for the shallower $^{23}$Na potential.  Nevertheless, the minima of the potentials are at the same position. (B) The effective trapping potential for $^{23}$Na with contributions from the interspecies interaction without (top) and with (bottom) gravity. The shaded areas show schematically the $^{23}$Na BEC filled in the shell-shaped trap for illustrating the origin of the asymmetric top and bottom atom distributions in the shell.  }
\label{figS1}
\end{figure}

The double BEC is first created in an optical dipole trap formed by two crossing 1070 nm laser beams. Details of the evaporative cooling in this step have been discussed before~\cite{Wang2013,Wang2015}. At this wavelength, the trap oscillation frequency of $^{23}$Na atoms is about 10\% higher than that of the $^{87}$Rb atoms. The differential gravity sag is thus non-zero and the centers of mass of the two BECs are displaced from each other in the vertical direction. To overcome this issue, we load the double BEC adiabatically into the 946 nm magic wavelength crossed ODT in which the two BECs experience the same trap frequencies~\cite{Safronova2006}. This is achieved by ramping off the 1070 nm light beams while ramping on two 946 nm beams in the same beam path. Figure~\ref{figS1}A shows the external trap potentials without (top) and with (bottom) the gravitational effects in the magic wavelength trap. The gravity sag shifts the minima of the potentials downward by the same amounts. As the horizontal trap potential are not affected by gravity, the two BECs produced in this trap will share the same center of mass positions.

The final configuration for studying the shell BEC is reached after further evaporative cooling by lowering the power of the 946 nm light. At the end of this procedure, two condensates with typically $1.3 \times 10^5$ $^{23}$Na atoms and $1.2\times 10^5$ $^{87}$Rb atoms without discernible thermal fractions are obtained routinely.  

An intuitive way to understand the shape of the $^{23}$Na shell is by looking at its effective trapping potential $V_1^{eff} = V_1 + U_{12} n_2$ as illustrated in Fig.~\ref{figS1}B without (top) and with (bottom) gravity. Without gravity, the repulsive interspecies interaction divides $V_1^{eff}$ in the vertical direction into two parts symmetrically. In this case, the $^{23}$Na BEC will form a spherical shell. The distortion caused by gravity breaks this symmetry and leads to more atoms at the bottom. However, a closed shell can still form, unless the number of $^{87}$Rb atoms is exceedingly large and the number of $^{23}$Na is really small. This picture agrees well with results obtained from numerical calculations of the coupled GP equations shown in Fig.~\ref{fig1} of the main text.

\subsection{Probing the shell BEC}

\begin{figure}[t]
\begin{center}
\includegraphics[width = 0.9 \linewidth]{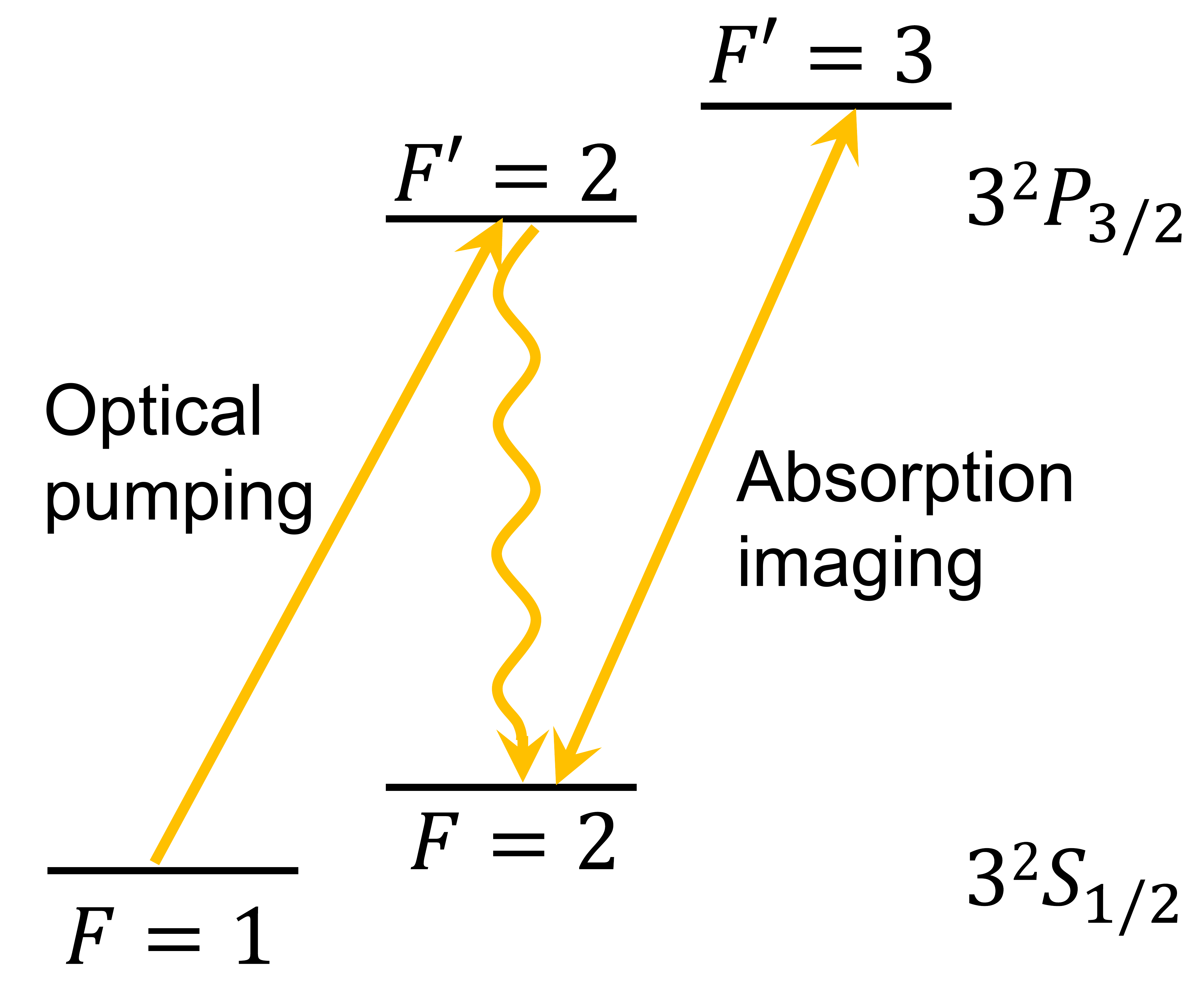}
\end{center}
\caption{Detection scheme of the $^{23}$Na shell BEC. The atoms are first transferred from $F = 1$ to $F = 2$ via optical pumping before being detected with absorption imaging on the $F = 2 \rightarrow F'= 3$ cycling transition. The optical pumping beam has a thickness of 15 $\mu$m for resolving the structures of the expanded shell.    
}
\label{figS2}
\end{figure}

The relevant atomic levels for the $^{23}$Na BEC detection are depicted in Fig.~\ref{figS2}. As illustrated in Fig.~\ref{fig1}, to resolve the shell structure, we shape the optical pumping beam to a $15 \mu m \times 800 \mu m$ light sheet with cylindrical lenses. This allow us to cut out only a slice of the shell to be observed by absorption imaging with the probe beam. 

The thickness of the slice can be reduced by using an optical pumping beam with large detuning, low power, and short pulse duration. However, due to the Gaussian beam profile, the minimum thickness is still more or less limited to the beam waist. As the result of this finite thickness and the curvature of the shell, the observed shell radius tends to be smaller than the real value. Ideally, to measure the shell size as accurate as possible, we should use the light sheet configuration which gives the thinnest slice. However, following the TOF expansion, the number of atoms can be optically pumped, and thus the signal, will be reduced. This is especially problematic for observing the interference fringes which already contain very small amounts of atoms. 

With the understanding to these limitations, in the experiment, we adjust the optical pumping parameters empirically to achieve enough signal but at the same time minimize the thickness of the optically pumped slice of the shell.  

Two imaging configurations are used in the experiment: 

\indent	1) At very short total TOFs, the sizes of the BECs are only several microns and the inner radius of the $^{23}$Na shell is smaller than the light sheet thickness. We use a $15\times$ magnification imaging system with measured optical resolutions of 1.8 $\mu$m and 2.4 $\mu$m for $^{23}$Na at 589 nm and $^{87}$Rb at 780 nm, respectively. As the OD is high, to avoid strong attenuation of the optical pumping beam, its frequency is 160 MHz detuned from the $F=1 \rightarrow F'=2$ transition. This ensures all atoms in the light sheet are pumped to $F=2$ with a uniform probability. A short pulse duration of 2 $\mu$s is used, and the power is varied from 0.07 mW to 3 mW.  The data in Fig.~\ref{fig2}A and Fig.~\ref{fig2}B are obtained with this imaging configuration. 

\indent	2) At longer TOFs, the size of the shell becomes too large for the $15\times$ imaging system. To increase the field of view, we change to a $3 \times$ magnification imaging system with an estimated optical resolution of 4 $\mu$m. As the OD becomes lower following the expansion, attenuation of the optical pumping beam is not significant. The optical pumping light frequency is thus tuned to the $F=1 \rightarrow F'=2$ resonance to increase the signal. The optical pumping power and pulse duration are set to below 0.1 mW and 2 $\mu$s, respectively. The data in Fig.~\ref{fig3}A and Fig.~\ref{fig4} are obtained with this configuration.    

\subsection{Numerical simulation}

The in-trap density of $^{23}$Na in Fig.~\ref{fig1}B and the release energies in Fig.~\ref{fig2}D are obtained by solving the coupled GPEs numerically. The external potential $V_i$ includes contributions from both the ODT and gravity. The Gaussian-shaped ODT trap can be well approximated by a harmonic function near its center. However, the gravity magnifies the anharmonicity, and distorts the potential. In the simulation, we represent $V_i$  with a polynomial up to the 4th order
\begin{widetext}
\begin{equation}
V_i = \frac{m_i}{2} (\omega_x^2 x^2 + \omega_z^2 z^2 ) + \frac{m_i}{2} \omega_y^2 (y + \Delta y)^2 + \beta (y + \Delta y)^3 + \gamma (y + \Delta y)^4,   
\label{eqs1}
\end{equation}
\end{widetext}
where the ratio $\beta /(m_i \omega_y^2/2 )\sim 0.06$ and $\gamma /(m_i \omega_y^2/2 ) \sim 0.001$ are obtained from fitting to the calculated full trapping potential based on measured trap frequencies and beam waists.   

\begin{figure}[t]
\begin{center}
\includegraphics[width = 0.9 \linewidth]{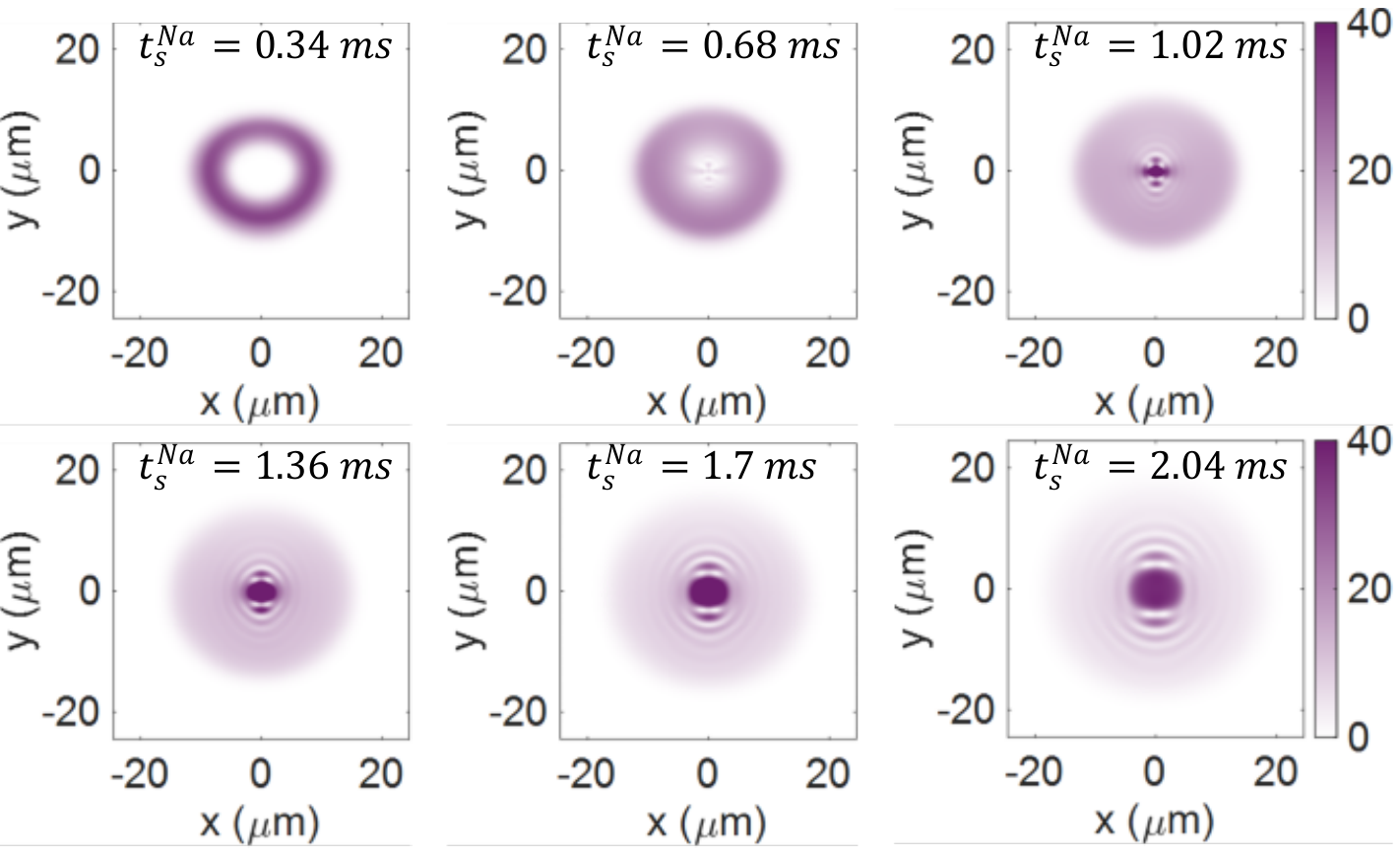}
\end{center}
\caption{Short time expansion dynamics of the hollow shell from numerical simulation. The shell is created after $t_{co} = 0.34$ ms. The implosion happens faster in the vertical direction due to the higher trap frequency. This also results in higher fringe contrast in the vertical direction. At 2.04 ms, the main features of the self-interference already match with those observed in the experiment. The increase in the outer size indicates explosion occurs simultaneously with implosion. The unit in the color bar is $\mu m^{-3}$.    
}
\label{figS3}
\end{figure}

Limited by the thickness resolution of the light sheet, the expansion dynamics of the hollow shell in very short time scales cannot be measured in the experiment. Figure~\ref{figS3} shows the dynamics in this period from numerical simulation for the hollow shell created after 0.34 ms co-expansion. In the first 2 ms, both implosion and explosion will happen. The self-interference fringes become visible once the atoms reach the center following the implosion at around 1 ms. Afterwards, both the number of atoms accumulated at the center and the fringe spacings increase with TOF. Importantly, due to the higher trap frequency in the vertical direction, the fringes in the x and y directions are not identical. In the vertical direction, the visibility is always higher. This is why we choose this part of the image in the data processing. The patterns at $t_s^{\rm Na}=2.04$ ms are already similar to the experimentally observed ones for longer $t_s^{\rm Na}$.

\subsection{Image processing and data analysis}

Due to the reduced number density, the signal of the shell decrease following the TOF expansion. For total TOF longer than 10 ms, the background noise, mainly because of light interference, becomes apparent. To increase the signal to noise ratio, we apply a fast Fourier transformation to the original OD image and mask out the high frequency noises. An OD image with improved quality is then obtained with an inverse fast Fourier transformation.     

To extract more detailed information of the shell and the interference fringes, we first convert the OD image from Cartesian coordinates to polar coordinates. For the expansion of the filled shell, the size is obtained from a Gaussian fitting to the azimuthally averaged 1D profile between polar angles $170^{\rm o}$ to $190^{\rm o}$.  The azimuthal average does not include the full $360^{\rm o}$ data as the image is not perfectly spherical. For the expansion of the hollow shell, to find the spacing between the self-interference fringes, we fit the data near each peak with a Gaussian to obtain the center of the peak. The fringe spacing is then the distance between the first and the second fringes.


\begin{thebibliography}{40}%
\makeatletter
\providecommand \@ifxundefined [1]{%
 \@ifx{#1\undefined}
}%
\providecommand \@ifnum [1]{%
 \ifnum #1\expandafter \@firstoftwo
 \else \expandafter \@secondoftwo
 \fi
}%
\providecommand \@ifx [1]{%
 \ifx #1\expandafter \@firstoftwo
 \else \expandafter \@secondoftwo
 \fi
}%
\providecommand \natexlab [1]{#1}%
\providecommand \enquote  [1]{``#1''}%
\providecommand \bibnamefont  [1]{#1}%
\providecommand \bibfnamefont [1]{#1}%
\providecommand \citenamefont [1]{#1}%
\providecommand \href@noop [0]{\@secondoftwo}%
\providecommand \href [0]{\begingroup \@sanitize@url \@href}%
\providecommand \@href[1]{\@@startlink{#1}\@@href}%
\providecommand \@@href[1]{\endgroup#1\@@endlink}%
\providecommand \@sanitize@url [0]{\catcode `\\12\catcode `\$12\catcode
  `\&12\catcode `\#12\catcode `\^12\catcode `\_12\catcode `\%12\relax}%
\providecommand \@@startlink[1]{}%
\providecommand \@@endlink[0]{}%
\providecommand \url  [0]{\begingroup\@sanitize@url \@url }%
\providecommand \@url [1]{\endgroup\@href {#1}{\urlprefix }}%
\providecommand \urlprefix  [0]{URL }%
\providecommand \Eprint [0]{\href }%
\providecommand \doibase [0]{https://doi.org/}%
\providecommand \selectlanguage [0]{\@gobble}%
\providecommand \bibinfo  [0]{\@secondoftwo}%
\providecommand \bibfield  [0]{\@secondoftwo}%
\providecommand \translation [1]{[#1]}%
\providecommand \BibitemOpen [0]{}%
\providecommand \bibitemStop [0]{}%
\providecommand \bibitemNoStop [0]{.\EOS\space}%
\providecommand \EOS [0]{\spacefactor3000\relax}%
\providecommand \BibitemShut  [1]{\csname bibitem#1\endcsname}%
\let\auto@bib@innerbib\@empty
%</preamble>
\bibitem [{\citenamefont {Bloch}\ \emph {et~al.}(2008)\citenamefont {Bloch},
  \citenamefont {Dalibard},\ and\ \citenamefont {Zwerger}}]{Bloch2008}%
  \BibitemOpen
  \bibfield  {author} {\bibinfo {author} {\bibfnamefont {I.}~\bibnamefont
  {Bloch}}, \bibinfo {author} {\bibfnamefont {J.}~\bibnamefont {Dalibard}},\
  and\ \bibinfo {author} {\bibfnamefont {W.}~\bibnamefont {Zwerger}},\
  }\bibfield  {title} {\bibinfo {title} {Many-body physics with ultracold
  gases},\ }\href {https://doi.org/10.1103/RevModPhys.80.885} {\bibfield
  {journal} {\bibinfo  {journal} {Rev. Mod. Phys.}\ }\textbf {\bibinfo {volume}
  {80}},\ \bibinfo {pages} {885} (\bibinfo {year} {2008})}\BibitemShut
  {NoStop}%
\bibitem [{\citenamefont {Greiner}\ \emph {et~al.}(2002)\citenamefont
  {Greiner}, \citenamefont {Mandel}, \citenamefont {Esslinger}, \citenamefont
  {H{\"a}nsch},\ and\ \citenamefont {Bloch}}]{Greiner2002}%
  \BibitemOpen
  \bibfield  {author} {\bibinfo {author} {\bibfnamefont {M.}~\bibnamefont
  {Greiner}}, \bibinfo {author} {\bibfnamefont {O.}~\bibnamefont {Mandel}},
  \bibinfo {author} {\bibfnamefont {T.}~\bibnamefont {Esslinger}}, \bibinfo
  {author} {\bibfnamefont {T.~W.}\ \bibnamefont {H{\"a}nsch}},\ and\ \bibinfo
  {author} {\bibfnamefont {I.}~\bibnamefont {Bloch}},\ }\bibfield  {title}
  {\bibinfo {title} {Quantum phase transition from a superfluid to a mott
  insulator in a gas of ultracold atoms},\ }\href@noop {} {\bibfield  {journal}
  {\bibinfo  {journal} {Nature}\ }\textbf {\bibinfo {volume} {415}},\ \bibinfo
  {pages} {39} (\bibinfo {year} {2002})}\BibitemShut {NoStop}%
\bibitem [{\citenamefont {Bloch}\ \emph {et~al.}(2012)\citenamefont {Bloch},
  \citenamefont {Dalibard},\ and\ \citenamefont {Nascimbene}}]{Bloch2012}%
  \BibitemOpen
  \bibfield  {author} {\bibinfo {author} {\bibfnamefont {I.}~\bibnamefont
  {Bloch}}, \bibinfo {author} {\bibfnamefont {J.}~\bibnamefont {Dalibard}},\
  and\ \bibinfo {author} {\bibfnamefont {S.}~\bibnamefont {Nascimbene}},\
  }\bibfield  {title} {\bibinfo {title} {Quantum simulations with ultracold
  quantum gases},\ }\href@noop {} {\bibfield  {journal} {\bibinfo  {journal}
  {Nat. Phys.}\ }\textbf {\bibinfo {volume} {8}},\ \bibinfo {pages} {267}
  (\bibinfo {year} {2012})}\BibitemShut {NoStop}%
\bibitem [{\citenamefont {Gross}\ and\ \citenamefont
  {Bloch}(2017)}]{Gross2017}%
  \BibitemOpen
  \bibfield  {author} {\bibinfo {author} {\bibfnamefont {C.}~\bibnamefont
  {Gross}}\ and\ \bibinfo {author} {\bibfnamefont {I.}~\bibnamefont {Bloch}},\
  }\bibfield  {title} {\bibinfo {title} {Quantum simulations with ultracold
  atoms in optical lattices},\ }\href {https://doi.org/10.1126/science.aal3837}
  {\bibfield  {journal} {\bibinfo  {journal} {Science}\ }\textbf {\bibinfo
  {volume} {357}},\ \bibinfo {pages} {995} (\bibinfo {year}
  {2017})}\BibitemShut {NoStop}%
\bibitem [{\citenamefont {Hadzibabic}\ and\ \citenamefont
  {Dalibard}(2011)}]{Hadzibabic2011}%
  \BibitemOpen
  \bibfield  {author} {\bibinfo {author} {\bibfnamefont {Z.}~\bibnamefont
  {Hadzibabic}}\ and\ \bibinfo {author} {\bibfnamefont {J.}~\bibnamefont
  {Dalibard}},\ }\bibfield  {title} {\bibinfo {title} {Two-dimensional bose
  fluids: An atomic physics perspective},\ }\href@noop {} {\bibfield  {journal}
  {\bibinfo  {journal} {Riv. Nuovo Cim.}\ }\textbf {\bibinfo {volume} {34}},\
  \bibinfo {pages} {389} (\bibinfo {year} {2011})}\BibitemShut {NoStop}%
\bibitem [{\citenamefont {Cazalilla}\ \emph {et~al.}(2011)\citenamefont
  {Cazalilla}, \citenamefont {Citro}, \citenamefont {Giamarchi}, \citenamefont
  {Orignac},\ and\ \citenamefont {Rigol}}]{Cazalilla2011}%
  \BibitemOpen
  \bibfield  {author} {\bibinfo {author} {\bibfnamefont {M.~A.}\ \bibnamefont
  {Cazalilla}}, \bibinfo {author} {\bibfnamefont {R.}~\bibnamefont {Citro}},
  \bibinfo {author} {\bibfnamefont {T.}~\bibnamefont {Giamarchi}}, \bibinfo
  {author} {\bibfnamefont {E.}~\bibnamefont {Orignac}},\ and\ \bibinfo {author}
  {\bibfnamefont {M.}~\bibnamefont {Rigol}},\ }\bibfield  {title} {\bibinfo
  {title} {One dimensional bosons: From condensed matter systems to ultracold
  gases},\ }\href {https://doi.org/10.1103/RevModPhys.83.1405} {\bibfield
  {journal} {\bibinfo  {journal} {Rev. Mod. Phys.}\ }\textbf {\bibinfo {volume}
  {83}},\ \bibinfo {pages} {1405} (\bibinfo {year} {2011})}\BibitemShut
  {NoStop}%
\bibitem [{\citenamefont {Navon}\ \emph {et~al.}(2021)\citenamefont {Navon},
  \citenamefont {Smith},\ and\ \citenamefont {Hadzibabic}}]{Navon2021}%
  \BibitemOpen
  \bibfield  {author} {\bibinfo {author} {\bibfnamefont {N.}~\bibnamefont
  {Navon}}, \bibinfo {author} {\bibfnamefont {R.~P.}\ \bibnamefont {Smith}},\
  and\ \bibinfo {author} {\bibfnamefont {Z.}~\bibnamefont {Hadzibabic}},\
  }\bibfield  {title} {\bibinfo {title} {Quantum gases in optical boxes},\
  }\href {https://doi.org/10.1038/s41567-021-01403-z} {\bibfield  {journal}
  {\bibinfo  {journal} {Nat. Phys.}\ }\textbf {\bibinfo {volume} {17}},\
  \bibinfo {pages} {334} (\bibinfo {year} {2021})}\BibitemShut {NoStop}%
\bibitem [{\citenamefont {Ramanathan}\ \emph {et~al.}(2011)\citenamefont
  {Ramanathan}, \citenamefont {Wright}, \citenamefont {Muniz}, \citenamefont
  {Zelan}, \citenamefont {Hill}, \citenamefont {Lobb}, \citenamefont
  {Helmerson}, \citenamefont {Phillips},\ and\ \citenamefont
  {Campbell}}]{Ramanathan2011}%
  \BibitemOpen
  \bibfield  {author} {\bibinfo {author} {\bibfnamefont {A.}~\bibnamefont
  {Ramanathan}}, \bibinfo {author} {\bibfnamefont {K.~C.}\ \bibnamefont
  {Wright}}, \bibinfo {author} {\bibfnamefont {S.~R.}\ \bibnamefont {Muniz}},
  \bibinfo {author} {\bibfnamefont {M.}~\bibnamefont {Zelan}}, \bibinfo
  {author} {\bibfnamefont {W.~T.}\ \bibnamefont {Hill}}, \bibinfo {author}
  {\bibfnamefont {C.~J.}\ \bibnamefont {Lobb}}, \bibinfo {author}
  {\bibfnamefont {K.}~\bibnamefont {Helmerson}}, \bibinfo {author}
  {\bibfnamefont {W.~D.}\ \bibnamefont {Phillips}},\ and\ \bibinfo {author}
  {\bibfnamefont {G.~K.}\ \bibnamefont {Campbell}},\ }\bibfield  {title}
  {\bibinfo {title} {Superflow in a toroidal bose-einstein condensate: An atom
  circuit with a tunable weak link},\ }\href
  {https://doi.org/10.1103/PhysRevLett.106.130401} {\bibfield  {journal}
  {\bibinfo  {journal} {Phys. Rev. Lett.}\ }\textbf {\bibinfo {volume} {106}},\
  \bibinfo {pages} {130401} (\bibinfo {year} {2011})}\BibitemShut {NoStop}%
\bibitem [{\citenamefont {Wright}\ \emph {et~al.}(2013)\citenamefont {Wright},
  \citenamefont {Blakestad}, \citenamefont {Lobb}, \citenamefont {Phillips},\
  and\ \citenamefont {Campbell}}]{Wright2013}%
  \BibitemOpen
  \bibfield  {author} {\bibinfo {author} {\bibfnamefont {K.~C.}\ \bibnamefont
  {Wright}}, \bibinfo {author} {\bibfnamefont {R.~B.}\ \bibnamefont
  {Blakestad}}, \bibinfo {author} {\bibfnamefont {C.~J.}\ \bibnamefont {Lobb}},
  \bibinfo {author} {\bibfnamefont {W.~D.}\ \bibnamefont {Phillips}},\ and\
  \bibinfo {author} {\bibfnamefont {G.~K.}\ \bibnamefont {Campbell}},\
  }\bibfield  {title} {\bibinfo {title} {Driving phase slips in a superfluid
  atom circuit with a rotating weak link},\ }\href
  {https://doi.org/10.1103/PhysRevLett.110.025302} {\bibfield  {journal}
  {\bibinfo  {journal} {Phys. Rev. Lett.}\ }\textbf {\bibinfo {volume} {110}},\
  \bibinfo {pages} {025302} (\bibinfo {year} {2013})}\BibitemShut {NoStop}%
\bibitem [{\citenamefont {Eckel}\ \emph {et~al.}(2018)\citenamefont {Eckel},
  \citenamefont {Kumar}, \citenamefont {Jacobson}, \citenamefont {Spielman},\
  and\ \citenamefont {Campbell}}]{Eckel2018}%
  \BibitemOpen
  \bibfield  {author} {\bibinfo {author} {\bibfnamefont {S.}~\bibnamefont
  {Eckel}}, \bibinfo {author} {\bibfnamefont {A.}~\bibnamefont {Kumar}},
  \bibinfo {author} {\bibfnamefont {T.}~\bibnamefont {Jacobson}}, \bibinfo
  {author} {\bibfnamefont {I.~B.}\ \bibnamefont {Spielman}},\ and\ \bibinfo
  {author} {\bibfnamefont {G.~K.}\ \bibnamefont {Campbell}},\ }\bibfield
  {title} {\bibinfo {title} {{A Rapidly Expanding Bose-Einstein Condensate: An
  Expanding Universe in the Lab}},\ }\href
  {https://doi.org/10.1103/PhysRevX.8.021021} {\bibfield  {journal} {\bibinfo
  {journal} {Phys. Rev. X}\ }\textbf {\bibinfo {volume} {8}},\ \bibinfo {pages}
  {021021} (\bibinfo {year} {2018})}\BibitemShut {NoStop}%
\bibitem [{\citenamefont {Zobay}\ and\ \citenamefont
  {Garraway}(2001)}]{Zobay2001}%
  \BibitemOpen
  \bibfield  {author} {\bibinfo {author} {\bibfnamefont {O.}~\bibnamefont
  {Zobay}}\ and\ \bibinfo {author} {\bibfnamefont {B.~M.}\ \bibnamefont
  {Garraway}},\ }\bibfield  {title} {\bibinfo {title} {{Two-dimensional atom
  trapping in field-induced adiabatic potentials}},\ }\href
  {https://doi.org/10.1103/PhysRevLett.86.1195} {\bibfield  {journal} {\bibinfo
   {journal} {Phys. Rev. Lett.}\ }\textbf {\bibinfo {volume} {86}},\ \bibinfo
  {pages} {1195} (\bibinfo {year} {2001})}\BibitemShut {NoStop}%
\bibitem [{\citenamefont {Lannert}\ \emph {et~al.}(2007)\citenamefont
  {Lannert}, \citenamefont {Wei},\ and\ \citenamefont
  {Vishveshwara}}]{Lannert2007}%
  \BibitemOpen
  \bibfield  {author} {\bibinfo {author} {\bibfnamefont {C.}~\bibnamefont
  {Lannert}}, \bibinfo {author} {\bibfnamefont {T.~C.}\ \bibnamefont {Wei}},\
  and\ \bibinfo {author} {\bibfnamefont {S.}~\bibnamefont {Vishveshwara}},\
  }\bibfield  {title} {\bibinfo {title} {{Dynamics of condensate shells:
  Collective modes and expansion}},\ }\href
  {https://doi.org/10.1103/PhysRevA.75.013611} {\bibfield  {journal} {\bibinfo
  {journal} {Phys. Rev. A}\ }\textbf {\bibinfo {volume} {75}},\ \bibinfo
  {pages} {1} (\bibinfo {year} {2007})}\BibitemShut {NoStop}%
\bibitem [{\citenamefont {Padavi{\'{c}}}\ \emph {et~al.}(2017)\citenamefont
  {Padavi{\'{c}}}, \citenamefont {Sun}, \citenamefont {Lannert},\ and\
  \citenamefont {Vishveshwara}}]{Padavic2017}%
  \BibitemOpen
  \bibfield  {author} {\bibinfo {author} {\bibfnamefont {K.}~\bibnamefont
  {Padavi{\'{c}}}}, \bibinfo {author} {\bibfnamefont {K.}~\bibnamefont {Sun}},
  \bibinfo {author} {\bibfnamefont {C.}~\bibnamefont {Lannert}},\ and\ \bibinfo
  {author} {\bibfnamefont {S.}~\bibnamefont {Vishveshwara}},\ }\bibfield
  {title} {\bibinfo {title} {{Physics of hollow Bose-Einstein condensates}},\
  }\href {https://doi.org/10.1209/0295-5075/120/20004} {\bibfield  {journal}
  {\bibinfo  {journal} {Euro. Lett.}\ }\textbf {\bibinfo {volume} {120}},\
  \bibinfo {pages} {20004} (\bibinfo {year} {2017})}\BibitemShut {NoStop}%
\bibitem [{\citenamefont {Sun}\ \emph {et~al.}(2018)\citenamefont {Sun},
  \citenamefont {Padavi{\'{c}}}, \citenamefont {Yang}, \citenamefont
  {Vishveshwara},\ and\ \citenamefont {Lannert}}]{Sun2018}%
  \BibitemOpen
  \bibfield  {author} {\bibinfo {author} {\bibfnamefont {K.}~\bibnamefont
  {Sun}}, \bibinfo {author} {\bibfnamefont {K.}~\bibnamefont {Padavi{\'{c}}}},
  \bibinfo {author} {\bibfnamefont {F.}~\bibnamefont {Yang}}, \bibinfo {author}
  {\bibfnamefont {S.}~\bibnamefont {Vishveshwara}},\ and\ \bibinfo {author}
  {\bibfnamefont {C.}~\bibnamefont {Lannert}},\ }\bibfield  {title} {\bibinfo
  {title} {{Static and dynamic properties of shell-shaped condensates}},\
  }\bibfield  {journal} {\bibinfo  {journal} {Phys. Rev. A}\ }\textbf {\bibinfo
  {volume} {98}},\ \href {https://doi.org/10.1103/PhysRevA.98.013609}
  {10.1103/PhysRevA.98.013609} (\bibinfo {year} {2018})\BibitemShut {NoStop}%
\bibitem [{\citenamefont {Tononi}\ and\ \citenamefont
  {Salasnich}(2019)}]{Tononi2019}%
  \BibitemOpen
  \bibfield  {author} {\bibinfo {author} {\bibfnamefont {A.}~\bibnamefont
  {Tononi}}\ and\ \bibinfo {author} {\bibfnamefont {L.}~\bibnamefont
  {Salasnich}},\ }\bibfield  {title} {\bibinfo {title} {{Bose-Einstein
  Condensation on the Surface of a Sphere}},\ }\href
  {https://doi.org/10.1103/PhysRevLett.123.160403} {\bibfield  {journal}
  {\bibinfo  {journal} {Phys. Rev. Lett.}\ }\textbf {\bibinfo {volume} {123}},\
  \bibinfo {pages} {160403} (\bibinfo {year} {2019})}\BibitemShut {NoStop}%
\bibitem [{\citenamefont {Tononi}\ \emph {et~al.}(2020)\citenamefont {Tononi},
  \citenamefont {Cinti},\ and\ \citenamefont {Salasnich}}]{Tononi2020}%
  \BibitemOpen
  \bibfield  {author} {\bibinfo {author} {\bibfnamefont {A.}~\bibnamefont
  {Tononi}}, \bibinfo {author} {\bibfnamefont {F.}~\bibnamefont {Cinti}},\ and\
  \bibinfo {author} {\bibfnamefont {L.}~\bibnamefont {Salasnich}},\ }\bibfield
  {title} {\bibinfo {title} {{Quantum Bubbles in Microgravity}},\ }\href
  {https://doi.org/10.1103/PhysRevLett.125.010402} {\bibfield  {journal}
  {\bibinfo  {journal} {Phys. Rev. Lett.}\ }\textbf {\bibinfo {volume} {125}},\
  \bibinfo {pages} {10402} (\bibinfo {year} {2020})}\BibitemShut {NoStop}%
\bibitem [{\citenamefont {Rhyno}\ \emph {et~al.}(2021)\citenamefont {Rhyno},
  \citenamefont {Lundblad}, \citenamefont {Aveline}, \citenamefont {Lannert},\
  and\ \citenamefont {Vishveshwara}}]{Rhyno2021}%
  \BibitemOpen
  \bibfield  {author} {\bibinfo {author} {\bibfnamefont {B.}~\bibnamefont
  {Rhyno}}, \bibinfo {author} {\bibfnamefont {N.}~\bibnamefont {Lundblad}},
  \bibinfo {author} {\bibfnamefont {D.~C.}\ \bibnamefont {Aveline}}, \bibinfo
  {author} {\bibfnamefont {C.}~\bibnamefont {Lannert}},\ and\ \bibinfo {author}
  {\bibfnamefont {S.}~\bibnamefont {Vishveshwara}},\ }\bibfield  {title}
  {\bibinfo {title} {Thermodynamics in expanding shell-shaped bose-einstein
  condensates},\ }\href {https://doi.org/10.1103/PhysRevA.104.063310}
  {\bibfield  {journal} {\bibinfo  {journal} {Phys. Rev. A}\ }\textbf {\bibinfo
  {volume} {104}},\ \bibinfo {pages} {063310} (\bibinfo {year}
  {2021})}\BibitemShut {NoStop}%
\bibitem [{\citenamefont {Padavi{\'{c}}}\ \emph {et~al.}(2020)\citenamefont
  {Padavi{\'{c}}}, \citenamefont {Sun}, \citenamefont {Lannert},\ and\
  \citenamefont {Vishveshwara}}]{Padavic2020}%
  \BibitemOpen
  \bibfield  {author} {\bibinfo {author} {\bibfnamefont {K.}~\bibnamefont
  {Padavi{\'{c}}}}, \bibinfo {author} {\bibfnamefont {K.}~\bibnamefont {Sun}},
  \bibinfo {author} {\bibfnamefont {C.}~\bibnamefont {Lannert}},\ and\ \bibinfo
  {author} {\bibfnamefont {S.}~\bibnamefont {Vishveshwara}},\ }\bibfield
  {title} {\bibinfo {title} {{Vortex-antivortex physics in shell-shaped
  Bose-Einstein condensates}},\ }\href
  {https://doi.org/10.1103/PhysRevA.102.043305} {\bibfield  {journal} {\bibinfo
   {journal} {Phys. Rev. A}\ }\textbf {\bibinfo {volume} {102}},\ \bibinfo
  {pages} {1} (\bibinfo {year} {2020})}\BibitemShut {NoStop}%
\bibitem [{\citenamefont {Zobay}\ and\ \citenamefont
  {Garraway}(2004)}]{Zobay2004}%
  \BibitemOpen
  \bibfield  {author} {\bibinfo {author} {\bibfnamefont {O.}~\bibnamefont
  {Zobay}}\ and\ \bibinfo {author} {\bibfnamefont {B.~M.}\ \bibnamefont
  {Garraway}},\ }\bibfield  {title} {\bibinfo {title} {{Atom trapping and
  two-dimensional Bose-Einstein condensates in field-induced adiabatic
  potentials}},\ }\href {https://doi.org/10.1103/PhysRevA.69.023605} {\bibfield
   {journal} {\bibinfo  {journal} {Phys. Rev. A}\ }\textbf {\bibinfo {volume}
  {69}},\ \bibinfo {pages} {15} (\bibinfo {year} {2004})}\BibitemShut {NoStop}%
\bibitem [{\citenamefont {Lundblad}\ \emph {et~al.}(2019)\citenamefont
  {Lundblad}, \citenamefont {Carollo}, \citenamefont {Lannert}, \citenamefont
  {Gold}, \citenamefont {Jiang}, \citenamefont {Paseltiner}, \citenamefont
  {Sergay},\ and\ \citenamefont {Aveline}}]{Lundblad2019}%
  \BibitemOpen
  \bibfield  {author} {\bibinfo {author} {\bibfnamefont {N.}~\bibnamefont
  {Lundblad}}, \bibinfo {author} {\bibfnamefont {R.~A.}\ \bibnamefont
  {Carollo}}, \bibinfo {author} {\bibfnamefont {C.}~\bibnamefont {Lannert}},
  \bibinfo {author} {\bibfnamefont {M.~J.}\ \bibnamefont {Gold}}, \bibinfo
  {author} {\bibfnamefont {X.}~\bibnamefont {Jiang}}, \bibinfo {author}
  {\bibfnamefont {D.}~\bibnamefont {Paseltiner}}, \bibinfo {author}
  {\bibfnamefont {N.}~\bibnamefont {Sergay}},\ and\ \bibinfo {author}
  {\bibfnamefont {D.~C.}\ \bibnamefont {Aveline}},\ }\bibfield  {title}
  {\bibinfo {title} {{Shell potentials for microgravity Bose–Einstein
  condensates}},\ }\href {https://doi.org/10.1038/s41526-019-0087-y} {\bibfield
   {journal} {\bibinfo  {journal} {npj Microgravity}\ }\textbf {\bibinfo
  {volume} {5}},\ \bibinfo {pages} {1} (\bibinfo {year} {2019})}\BibitemShut
  {NoStop}%
\bibitem [{\citenamefont {Carollo}\ \emph {et~al.}(2022)\citenamefont
  {Carollo}, \citenamefont {Aveline}, \citenamefont {Rhyno}, \citenamefont
  {Vishveshwara}, \citenamefont {Lannert}, \citenamefont {Murphree},
  \citenamefont {Elliott}, \citenamefont {Williams}, \citenamefont {Thompson},\
  and\ \citenamefont {Lundblad}}]{Carollo2021}%
  \BibitemOpen
  \bibfield  {author} {\bibinfo {author} {\bibfnamefont {R.~A.}\ \bibnamefont
  {Carollo}}, \bibinfo {author} {\bibfnamefont {D.~C.}\ \bibnamefont
  {Aveline}}, \bibinfo {author} {\bibfnamefont {B.}~\bibnamefont {Rhyno}},
  \bibinfo {author} {\bibfnamefont {S.}~\bibnamefont {Vishveshwara}}, \bibinfo
  {author} {\bibfnamefont {C.}~\bibnamefont {Lannert}}, \bibinfo {author}
  {\bibfnamefont {J.~D.}\ \bibnamefont {Murphree}}, \bibinfo {author}
  {\bibfnamefont {E.~R.}\ \bibnamefont {Elliott}}, \bibinfo {author}
  {\bibfnamefont {J.~R.}\ \bibnamefont {Williams}}, \bibinfo {author}
  {\bibfnamefont {R.~J.}\ \bibnamefont {Thompson}},\ and\ \bibinfo {author}
  {\bibfnamefont {N.}~\bibnamefont {Lundblad}},\ }\bibfield  {title} {\bibinfo
  {title} {Observation of ultracold atomic bubbles in orbital microgravity},\
  }\href {https://doi.org/10.1038/s41586-022-04639-8} {\bibfield  {journal}
  {\bibinfo  {journal} {Nature}\ }\textbf {\bibinfo {volume} {606}},\ \bibinfo
  {pages} {pages 281} (\bibinfo {year} {2022})}\BibitemShut {NoStop}%
\bibitem [{\citenamefont {Ho}\ and\ \citenamefont {Shenoy}(1996)}]{Ho1996}%
  \BibitemOpen
  \bibfield  {author} {\bibinfo {author} {\bibfnamefont {T.-L.}\ \bibnamefont
  {Ho}}\ and\ \bibinfo {author} {\bibfnamefont {V.~B.}\ \bibnamefont
  {Shenoy}},\ }\bibfield  {title} {\bibinfo {title} {{Binary Mixtures of Bose
  Condensates of Alkali Atoms}},\ }\href
  {https://doi.org/10.1103/PhysRevLett.77.3276} {\bibfield  {journal} {\bibinfo
   {journal} {Phys. Rev. Lett.}\ }\textbf {\bibinfo {volume} {77}},\ \bibinfo
  {pages} {3276} (\bibinfo {year} {1996})}\BibitemShut {NoStop}%
\bibitem [{\citenamefont {Pu}\ and\ \citenamefont {Bigelow}(1998)}]{Pu1998}%
  \BibitemOpen
  \bibfield  {author} {\bibinfo {author} {\bibfnamefont {H.}~\bibnamefont
  {Pu}}\ and\ \bibinfo {author} {\bibfnamefont {N.~P.}\ \bibnamefont
  {Bigelow}},\ }\bibfield  {title} {\bibinfo {title} {{Properties of
  two-species Bose condensates}},\ }\href
  {https://doi.org/10.1103/PhysRevLett.80.1130} {\bibfield  {journal} {\bibinfo
   {journal} {Phys. Rev. Lett.}\ }\textbf {\bibinfo {volume} {80}},\ \bibinfo
  {pages} {1130} (\bibinfo {year} {1998})}\BibitemShut {NoStop}%
\bibitem [{\citenamefont {Trippenbach}\ \emph {et~al.}(2000)\citenamefont
  {Trippenbach}, \citenamefont {G{\'{o}}ral}, \citenamefont {Rzazewski},
  \citenamefont {Malomed},\ and\ \citenamefont {Band}}]{Trippenbach2000}%
  \BibitemOpen
  \bibfield  {author} {\bibinfo {author} {\bibfnamefont {M.}~\bibnamefont
  {Trippenbach}}, \bibinfo {author} {\bibfnamefont {K.}~\bibnamefont
  {G{\'{o}}ral}}, \bibinfo {author} {\bibfnamefont {K.}~\bibnamefont
  {Rzazewski}}, \bibinfo {author} {\bibfnamefont {B.}~\bibnamefont {Malomed}},\
  and\ \bibinfo {author} {\bibfnamefont {Y.~B.}\ \bibnamefont {Band}},\
  }\bibfield  {title} {\bibinfo {title} {{Structure of binary Bose-Einstein
  condensates}},\ }\href {https://doi.org/10.1088/0953-4075/33/19/314}
  {\bibfield  {journal} {\bibinfo  {journal} {J. Phys. B}\ }\textbf {\bibinfo
  {volume} {33}},\ \bibinfo {pages} {4017} (\bibinfo {year}
  {2000})}\BibitemShut {NoStop}%
\bibitem [{\citenamefont {Lee}\ \emph {et~al.}(2016)\citenamefont {Lee},
  \citenamefont {J{\o}rgensen}, \citenamefont {Liu}, \citenamefont {Wacker},
  \citenamefont {Arlt},\ and\ \citenamefont {Proukakis}}]{Lee2016}%
  \BibitemOpen
  \bibfield  {author} {\bibinfo {author} {\bibfnamefont {K.~L.}\ \bibnamefont
  {Lee}}, \bibinfo {author} {\bibfnamefont {N.~B.}\ \bibnamefont
  {J{\o}rgensen}}, \bibinfo {author} {\bibfnamefont {I.~K.}\ \bibnamefont
  {Liu}}, \bibinfo {author} {\bibfnamefont {L.}~\bibnamefont {Wacker}},
  \bibinfo {author} {\bibfnamefont {J.~J.}\ \bibnamefont {Arlt}},\ and\
  \bibinfo {author} {\bibfnamefont {N.~P.}\ \bibnamefont {Proukakis}},\
  }\bibfield  {title} {\bibinfo {title} {{Phase separation and dynamics of
  two-component Bose-Einstein condensates}},\ }\href
  {https://doi.org/10.1103/PhysRevA.94.013602} {\bibfield  {journal} {\bibinfo
  {journal} {Phys. Rev. A}\ }\textbf {\bibinfo {volume} {94}},\ \bibinfo
  {pages} {1} (\bibinfo {year} {2016})}\BibitemShut {NoStop}%
\bibitem [{\citenamefont {Wolf}\ \emph {et~al.}(2022)\citenamefont {Wolf},
  \citenamefont {Boegel}, \citenamefont {Meister}, \citenamefont
  {Bala\ifmmode~\check{z}\else \v{z}\fi{}}, \citenamefont {Gaaloul},\ and\
  \citenamefont {Efremov}}]{Wolf2021}%
  \BibitemOpen
  \bibfield  {author} {\bibinfo {author} {\bibfnamefont {A.}~\bibnamefont
  {Wolf}}, \bibinfo {author} {\bibfnamefont {P.}~\bibnamefont {Boegel}},
  \bibinfo {author} {\bibfnamefont {M.}~\bibnamefont {Meister}}, \bibinfo
  {author} {\bibfnamefont {A.}~\bibnamefont {Bala\ifmmode~\check{z}\else
  \v{z}\fi{}}}, \bibinfo {author} {\bibfnamefont {N.}~\bibnamefont {Gaaloul}},\
  and\ \bibinfo {author} {\bibfnamefont {M.~A.}\ \bibnamefont {Efremov}},\
  }\bibfield  {title} {\bibinfo {title} {{Shell-shaped Bose-Einstein
  condensates based on dual-species mixtures}},\ }\href
  {https://doi.org/10.1103/PhysRevA.106.013309} {\bibfield  {journal} {\bibinfo
   {journal} {Phys. Rev. A}\ }\textbf {\bibinfo {volume} {106}},\ \bibinfo
  {pages} {013309} (\bibinfo {year} {2022})}\BibitemShut {NoStop}%
\bibitem [{\citenamefont {Wang}\ \emph {et~al.}(2013)\citenamefont {Wang},
  \citenamefont {Xiong}, \citenamefont {Li}, \citenamefont {Wang},\ and\
  \citenamefont {Tiemann}}]{Wang2013}%
  \BibitemOpen
  \bibfield  {author} {\bibinfo {author} {\bibfnamefont {F.}~\bibnamefont
  {Wang}}, \bibinfo {author} {\bibfnamefont {D.}~\bibnamefont {Xiong}},
  \bibinfo {author} {\bibfnamefont {X.}~\bibnamefont {Li}}, \bibinfo {author}
  {\bibfnamefont {D.}~\bibnamefont {Wang}},\ and\ \bibinfo {author}
  {\bibfnamefont {E.}~\bibnamefont {Tiemann}},\ }\bibfield  {title} {\bibinfo
  {title} {{Observation of Feshbach resonances between ultracold Na and Rb
  atoms}},\ }\href {https://doi.org/10.1103/PhysRevA.87.050702} {\bibfield
  {journal} {\bibinfo  {journal} {Phys. Rev. A}\ }\textbf {\bibinfo {volume}
  {87}},\ \bibinfo {pages} {1} (\bibinfo {year} {2013})}\BibitemShut {NoStop}%
\bibitem [{\citenamefont {Wang}\ \emph {et~al.}(2015)\citenamefont {Wang},
  \citenamefont {Li}, \citenamefont {Xiong},\ and\ \citenamefont
  {Wang}}]{Wang2015}%
  \BibitemOpen
  \bibfield  {author} {\bibinfo {author} {\bibfnamefont {F.}~\bibnamefont
  {Wang}}, \bibinfo {author} {\bibfnamefont {X.}~\bibnamefont {Li}}, \bibinfo
  {author} {\bibfnamefont {D.}~\bibnamefont {Xiong}},\ and\ \bibinfo {author}
  {\bibfnamefont {D.}~\bibnamefont {Wang}},\ }\bibfield  {title} {\bibinfo
  {title} {{A double species $^{23}$Na and $^{87}$Rb Bose-Einstein condensate
  with tunable miscibility via an interspecies Feshbach resonance}},\ }\href
  {https://doi.org/10.1088/0953-4075/49/1/015302} {\bibfield  {journal}
  {\bibinfo  {journal} {J. Phys. B}\ }\textbf {\bibinfo {volume} {49}},\
  \bibinfo {pages} {015302} (\bibinfo {year} {2015})}\BibitemShut {NoStop}%
\bibitem [{\citenamefont {Safronova}\ \emph {et~al.}(2006)\citenamefont
  {Safronova}, \citenamefont {Arora},\ and\ \citenamefont
  {Clark}}]{Safronova2006}%
  \BibitemOpen
  \bibfield  {author} {\bibinfo {author} {\bibfnamefont {M.~S.}\ \bibnamefont
  {Safronova}}, \bibinfo {author} {\bibfnamefont {B.}~\bibnamefont {Arora}},\
  and\ \bibinfo {author} {\bibfnamefont {C.~W.}\ \bibnamefont {Clark}},\
  }\bibfield  {title} {\bibinfo {title} {{Frequency-dependent polarizabilities
  of alkali-metal atoms from ultraviolet through infrared spectral regions}},\
  }\href {https://doi.org/10.1103/PhysRevA.73.022505} {\bibfield  {journal}
  {\bibinfo  {journal} {Phys. Rev. A}\ }\textbf {\bibinfo {volume} {73}},\
  \bibinfo {pages} {1} (\bibinfo {year} {2006})}\BibitemShut {NoStop}%
\bibitem [{\citenamefont {Knoop}\ \emph {et~al.}(2011)\citenamefont {Knoop},
  \citenamefont {Schuster}, \citenamefont {Scelle}, \citenamefont {Trautmann},
  \citenamefont {Appmeier}, \citenamefont {Oberthaler}, \citenamefont
  {Tiesinga},\ and\ \citenamefont {Tiemann}}]{Knoop2011}%
  \BibitemOpen
  \bibfield  {author} {\bibinfo {author} {\bibfnamefont {S.}~\bibnamefont
  {Knoop}}, \bibinfo {author} {\bibfnamefont {T.}~\bibnamefont {Schuster}},
  \bibinfo {author} {\bibfnamefont {R.}~\bibnamefont {Scelle}}, \bibinfo
  {author} {\bibfnamefont {A.}~\bibnamefont {Trautmann}}, \bibinfo {author}
  {\bibfnamefont {J.}~\bibnamefont {Appmeier}}, \bibinfo {author}
  {\bibfnamefont {M.~K.}\ \bibnamefont {Oberthaler}}, \bibinfo {author}
  {\bibfnamefont {E.}~\bibnamefont {Tiesinga}},\ and\ \bibinfo {author}
  {\bibfnamefont {E.}~\bibnamefont {Tiemann}},\ }\bibfield  {title} {\bibinfo
  {title} {Feshbach spectroscopy and analysis of the interaction potentials of
  ultracold sodium},\ }\href {https://doi.org/10.1103/PhysRevA.83.042704}
  {\bibfield  {journal} {\bibinfo  {journal} {Phys. Rev. A}\ }\textbf {\bibinfo
  {volume} {83}},\ \bibinfo {pages} {042704} (\bibinfo {year}
  {2011})}\BibitemShut {NoStop}%
\bibitem [{\citenamefont {van Kempen}\ \emph {et~al.}(2002)\citenamefont {van
  Kempen}, \citenamefont {Kokkelmans}, \citenamefont {Heinzen},\ and\
  \citenamefont {Verhaar}}]{Kempen2002}%
  \BibitemOpen
  \bibfield  {author} {\bibinfo {author} {\bibfnamefont {E.~G.~M.}\
  \bibnamefont {van Kempen}}, \bibinfo {author} {\bibfnamefont {S.~J. J.
  M.~F.}\ \bibnamefont {Kokkelmans}}, \bibinfo {author} {\bibfnamefont {D.~J.}\
  \bibnamefont {Heinzen}},\ and\ \bibinfo {author} {\bibfnamefont {B.~J.}\
  \bibnamefont {Verhaar}},\ }\bibfield  {title} {\bibinfo {title} {Interisotope
  determination of ultracold rubidium interactions from three high-precision
  experiments},\ }\href {https://doi.org/10.1103/PhysRevLett.88.093201}
  {\bibfield  {journal} {\bibinfo  {journal} {Phys. Rev. Lett.}\ }\textbf
  {\bibinfo {volume} {88}},\ \bibinfo {pages} {093201} (\bibinfo {year}
  {2002})}\BibitemShut {NoStop}%
\bibitem [{\citenamefont {Guo}\ \emph {et~al.}(2022)\citenamefont {Guo},
  \citenamefont {Jia}, \citenamefont {Zhu}, \citenamefont {Li}, \citenamefont
  {Hutson},\ and\ \citenamefont {Wang}}]{Guo2022}%
  \BibitemOpen
  \bibfield  {author} {\bibinfo {author} {\bibfnamefont {Z.}~\bibnamefont
  {Guo}}, \bibinfo {author} {\bibfnamefont {F.}~\bibnamefont {Jia}}, \bibinfo
  {author} {\bibfnamefont {B.}~\bibnamefont {Zhu}}, \bibinfo {author}
  {\bibfnamefont {L.}~\bibnamefont {Li}}, \bibinfo {author} {\bibfnamefont
  {J.~M.}\ \bibnamefont {Hutson}},\ and\ \bibinfo {author} {\bibfnamefont
  {D.}~\bibnamefont {Wang}},\ }\bibfield  {title} {\bibinfo {title} {Improved
  characterization of feshbach resonances and interaction potentials between
  $^{23}\mathrm{Na}$ and $^{87}\mathrm{Rb}$ atoms},\ }\href
  {https://doi.org/10.1103/PhysRevA.105.023313} {\bibfield  {journal} {\bibinfo
   {journal} {Phys. Rev. A}\ }\textbf {\bibinfo {volume} {105}},\ \bibinfo
  {pages} {023313} (\bibinfo {year} {2022})}\BibitemShut {NoStop}%
\bibitem [{Not()}]{Note1}%
  \BibitemOpen
  \href@noop {} {}\bibinfo {note} {Materials and methods are available as
  supplementary materials.}\BibitemShut {Stop}%
\bibitem [{\citenamefont {Antoine}\ and\ \citenamefont
  {Duboscq}(2014)}]{Antoine2014}%
  \BibitemOpen
  \bibfield  {author} {\bibinfo {author} {\bibfnamefont {X.}~\bibnamefont
  {Antoine}}\ and\ \bibinfo {author} {\bibfnamefont {R.}~\bibnamefont
  {Duboscq}},\ }\bibfield  {title} {\bibinfo {title} {{GPELab, a Matlab toolbox
  to solve Gross-Pitaevskii equations I: Computation of stationary
  solutions}},\ }\href {https://doi.org/10.1016/j.cpc.2014.06.026} {\bibfield
  {journal} {\bibinfo  {journal} {Comput. Phys. Commun.}\ }\textbf {\bibinfo
  {volume} {185}},\ \bibinfo {pages} {2969} (\bibinfo {year}
  {2014})}\BibitemShut {NoStop}%
\bibitem [{\citenamefont {Andrews}\ \emph {et~al.}(1997)\citenamefont
  {Andrews}, \citenamefont {Townsend}, \citenamefont {Miesner}, \citenamefont
  {Durfee}, \citenamefont {Kurn},\ and\ \citenamefont
  {Ketterle}}]{Andrews1997}%
  \BibitemOpen
  \bibfield  {author} {\bibinfo {author} {\bibfnamefont {M.~R.}\ \bibnamefont
  {Andrews}}, \bibinfo {author} {\bibfnamefont {C.~G.}\ \bibnamefont
  {Townsend}}, \bibinfo {author} {\bibfnamefont {H.-J.}\ \bibnamefont
  {Miesner}}, \bibinfo {author} {\bibfnamefont {D.~S.}\ \bibnamefont {Durfee}},
  \bibinfo {author} {\bibfnamefont {D.~M.}\ \bibnamefont {Kurn}},\ and\
  \bibinfo {author} {\bibfnamefont {W.}~\bibnamefont {Ketterle}},\ }\bibfield
  {title} {\bibinfo {title} {{Observation of Interference Between Two Bose
  Condensates}},\ }\href {https://doi.org/10.1126/science.275.5300.637}
  {\bibfield  {journal} {\bibinfo  {journal} {Science}\ }\textbf {\bibinfo
  {volume} {275}},\ \bibinfo {pages} {637} (\bibinfo {year}
  {1997})}\BibitemShut {NoStop}%
\bibitem [{\citenamefont {Mewes}\ \emph {et~al.}(1996)\citenamefont {Mewes},
  \citenamefont {Andrews}, \citenamefont {van Druten}, \citenamefont {Kurn},
  \citenamefont {Durfee},\ and\ \citenamefont {Ketterle}}]{Mewes1996}%
  \BibitemOpen
  \bibfield  {author} {\bibinfo {author} {\bibfnamefont {M.-O.}\ \bibnamefont
  {Mewes}}, \bibinfo {author} {\bibfnamefont {M.~R.}\ \bibnamefont {Andrews}},
  \bibinfo {author} {\bibfnamefont {N.~J.}\ \bibnamefont {van Druten}},
  \bibinfo {author} {\bibfnamefont {D.~M.}\ \bibnamefont {Kurn}}, \bibinfo
  {author} {\bibfnamefont {D.~S.}\ \bibnamefont {Durfee}},\ and\ \bibinfo
  {author} {\bibfnamefont {W.}~\bibnamefont {Ketterle}},\ }\bibfield  {title}
  {\bibinfo {title} {{Bose-Einstein} condensation in a tightly confining dc
  magnetic trap},\ }\href {https://doi.org/10.1103/PhysRevLett.77.416}
  {\bibfield  {journal} {\bibinfo  {journal} {Phys. Rev. Lett.}\ }\textbf
  {\bibinfo {volume} {77}},\ \bibinfo {pages} {416} (\bibinfo {year}
  {1996})}\BibitemShut {NoStop}%
\bibitem [{\citenamefont {Holland}\ \emph {et~al.}(1997)\citenamefont
  {Holland}, \citenamefont {Jin}, \citenamefont {Chiofalo},\ and\ \citenamefont
  {Cooper}}]{Holland1997}%
  \BibitemOpen
  \bibfield  {author} {\bibinfo {author} {\bibfnamefont {M.~J.}\ \bibnamefont
  {Holland}}, \bibinfo {author} {\bibfnamefont {D.~S.}\ \bibnamefont {Jin}},
  \bibinfo {author} {\bibfnamefont {M.~L.}\ \bibnamefont {Chiofalo}},\ and\
  \bibinfo {author} {\bibfnamefont {J.}~\bibnamefont {Cooper}},\ }\bibfield
  {title} {\bibinfo {title} {Emergence of interaction effects in
  {Bose-Einstein} condensation},\ }\href
  {https://doi.org/10.1103/PhysRevLett.78.3801} {\bibfield  {journal} {\bibinfo
   {journal} {Phys. Rev. Lett.}\ }\textbf {\bibinfo {volume} {78}},\ \bibinfo
  {pages} {3801} (\bibinfo {year} {1997})}\BibitemShut {NoStop}%
\bibitem [{\citenamefont {R\"ohrl}\ \emph {et~al.}(1997)\citenamefont
  {R\"ohrl}, \citenamefont {Naraschewski}, \citenamefont {Schenzle},\ and\
  \citenamefont {Wallis}}]{Rohrl1997}%
  \BibitemOpen
  \bibfield  {author} {\bibinfo {author} {\bibfnamefont {A.}~\bibnamefont
  {R\"ohrl}}, \bibinfo {author} {\bibfnamefont {M.}~\bibnamefont
  {Naraschewski}}, \bibinfo {author} {\bibfnamefont {A.}~\bibnamefont
  {Schenzle}},\ and\ \bibinfo {author} {\bibfnamefont {H.}~\bibnamefont
  {Wallis}},\ }\bibfield  {title} {\bibinfo {title} {{Transition from Phase
  Locking to the Interference of Independent Bose Condensates: Theory versus
  Experiment}},\ }\href {https://doi.org/10.1103/PhysRevLett.78.4143}
  {\bibfield  {journal} {\bibinfo  {journal} {Phys. Rev. Lett.}\ }\textbf
  {\bibinfo {volume} {78}},\ \bibinfo {pages} {4143} (\bibinfo {year}
  {1997})}\BibitemShut {NoStop}%
\bibitem [{\citenamefont {Dalfovo}\ \emph {et~al.}(1999)\citenamefont
  {Dalfovo}, \citenamefont {Giorgini}, \citenamefont {Pitaevskii},\ and\
  \citenamefont {Stringari}}]{Dalfovo1999}%
  \BibitemOpen
  \bibfield  {author} {\bibinfo {author} {\bibfnamefont {F.}~\bibnamefont
  {Dalfovo}}, \bibinfo {author} {\bibfnamefont {S.}~\bibnamefont {Giorgini}},
  \bibinfo {author} {\bibfnamefont {L.~P.}\ \bibnamefont {Pitaevskii}},\ and\
  \bibinfo {author} {\bibfnamefont {S.}~\bibnamefont {Stringari}},\ }\bibfield
  {title} {\bibinfo {title} {{Theory of Bose-Einstein condensation in trapped
  gases}},\ }\href {https://doi.org/10.1103/RevModPhys.71.463} {\bibfield
  {journal} {\bibinfo  {journal} {Rev. Mod. Phys.}\ }\textbf {\bibinfo {volume}
  {71}},\ \bibinfo {pages} {463} (\bibinfo {year} {1999})}\BibitemShut
  {NoStop}%
\bibitem [{\citenamefont {Guo}\ \emph {et~al.}(2021)\citenamefont {Guo},
  \citenamefont {Jia}, \citenamefont {Li}, \citenamefont {Ma}, \citenamefont
  {Hutson}, \citenamefont {Cui},\ and\ \citenamefont {Wang}}]{Guo2021}%
  \BibitemOpen
  \bibfield  {author} {\bibinfo {author} {\bibfnamefont {Z.}~\bibnamefont
  {Guo}}, \bibinfo {author} {\bibfnamefont {F.}~\bibnamefont {Jia}}, \bibinfo
  {author} {\bibfnamefont {L.}~\bibnamefont {Li}}, \bibinfo {author}
  {\bibfnamefont {Y.}~\bibnamefont {Ma}}, \bibinfo {author} {\bibfnamefont
  {J.~M.}\ \bibnamefont {Hutson}}, \bibinfo {author} {\bibfnamefont
  {X.}~\bibnamefont {Cui}},\ and\ \bibinfo {author} {\bibfnamefont
  {D.}~\bibnamefont {Wang}},\ }\bibfield  {title} {\bibinfo {title}
  {{Lee-Huang-Yang effects in the ultracold mixture of $^{23}$Na and $^{87}$Rb
  with attractive interspecies interactions}},\ }\href
  {https://doi.org/10.1103/PhysRevResearch.3.033247} {\bibfield  {journal}
  {\bibinfo  {journal} {Phys. Rev. Research}\ }\textbf {\bibinfo {volume}
  {3}},\ \bibinfo {pages} {1} (\bibinfo {year} {2021})}\BibitemShut {NoStop}%
\end{thebibliography}
\end{document}